\documentclass[12pt]{article}
\input epsf
\usepackage{epsfig
}

\def\IZ{\relax\ifmmode\mathchoice
{\hbox{\cmss Z\kern-.4em Z}}{\hbox{\cmss Z\kern-.4em Z}}
{\lower.9pt\hbox{\cmsss Z\kern-.4em Z}} {\lower1.2pt\hbox{\cmsss
Z\kern-.4em Z}}\else{\cmss Z\kern-.4em Z}\fi}
\def\IR{\relax{\rm I\kern-.18em R}}

\def\one{{\hbox{ 1\kern-.8mm l}}}

\newcommand{\N}{{\cal N}}
\newlength{\bredde}
\def\slash#1{\settowidth{\bredde}{$#1$}\ifmmode\,\raisebox{.15ex}{/}
\hspace*{-\bredde} #1\else$\,\raisebox{.15ex}{/}\hspace*{-\bredde}
#1$\fi}

\newcommand  {\Rbar} {{\mbox{\rm$\mbox{I}\!\mbox{R}$}}}

\newsavebox{\zzzbar}
\sbox{\zzzbar}
  {\setlength{\unitlength}{0.9em}
  \begin{picture}(0.6,0.7)
  \thinlines
  \put(0,0){\line(1,0){0.6}}
  \put(0,0.75){\line(1,0){0.575}}
  \multiput(0,0)(0.0125,0.025){30}{\rule{0.3pt}{0.3pt}}
  \multiput(0.2,0)(0.0125,0.025){30}{\rule{0.3pt}{0.3pt}}
  \put(0,0.75){\line(0,-1){0.15}}
  \put(0.015,0.75){\line(0,-1){0.1}}
  \put(0.03,0.75){\line(0,-1){0.075}}
  \put(0.045,0.75){\line(0,-1){0.05}}
  \put(0.05,0.75){\line(0,-1){0.025}}
  \put(0.6,0){\line(0,1){0.15}}
  \put(0.585,0){\line(0,1){0.1}}
  \put(0.57,0){\line(0,1){0.075}}
  \put(0.555,0){\line(0,1){0.05}}
  \put(0.55,0){\line(0,1){0.025}}
  \end{picture}}
\newcommand{\Zbar}{\mathord{\!{\usebox{\zzzbar}}}}

\newcommand{\ena}{\end{eqnarray}}
\newcommand{\beqa}{\begin{eqnarray}}
\newcommand{\eeqa}{\end{eqnarray}}

\newcommand{\eq}[1]{(\ref{#1})}
\newcommand{\fig}[1]{Fig.~\ref{#1}}

\newcommand{\be}{\begin{equation}}
\newcommand{\ee}{\end{equation}}
\def\G{\Gamma}
\def\K{{\cal K}}
\def\N{{\cal N}}
\def\H{{\cal H}}
\begin{document}

\begin{titlepage}
\begin{flushright}
EFI-03-30 \\
hep-th/0308057
\end{flushright}
\begin{center}
{\LARGE\bf Global Fluctuation Spectra \\ \vskip 3mm in Big Crunch/Big Bang String Vacua}    \\
\vskip 10.mm
{  Ben Craps $^{1}$ and Burt A.~Ovrut $^2$ } \\
\vskip 7mm
{\em $^1$ Enrico Fermi Institute, University of Chicago, 5640 S.~Ellis Av., Chicago, IL 60637, USA}\\
\vskip 0.5cm
{\em $^2$ Department of Physics, University of Pennsylvania, Philadelphia, PA 19104--6396}
\end{center}
\vfill

\begin{center}
{\bf ABSTRACT}
\end{center}
We study Big Crunch/Big Bang cosmologies that correspond to exact world-sheet
superconformal field theories of type II strings. The string
theory spacetime contains a Big Crunch and a Big Bang cosmology,
as well as additional ``whisker'' asymptotic and intermediate regions.
Within the context of free string theory, we compute,
unambiguously, the scalar fluctuation spectrum in all regions of spacetime. Generically,
the Big Crunch fluctuation spectrum is altered
while passing through the bounce singularity. The change in the spectrum is
characterized by a function $\Delta$, which is momentum and time-dependent.
We compute $\Delta$ explicitly and
demonstrate that it arises from the whisker regions.
The whiskers are also shown to lead to ``entanglement'' entropy in
the Big Bang region.
Finally, in the Milne orbifold limit of our superconformal vacua, we show that
$\Delta\rightarrow 1$ and, hence, the fluctuation spectrum is unaltered by the
Big Crunch/Big Bang singularity. We comment on, but do not attempt
to resolve, subtleties related to gravitational backreaction and
light winding modes when interactions are taken into account.
\vfill
\hrule width 5.cm
\vskip 2.mm
{\small
\noindent  {\tt craps@theory.uchicago.edu,  ovrut@ovrut.hep.upenn.edu}}
\end{titlepage}

\section{Introduction}

The theory of Ekpyrotic cosmology~\cite{cosmoA,cosmoB,cosmoC}, as well as
pre Big Bang scenarios~\cite{Gasperini:2002bn}, has emphasized the idea that the Universe
did not begin at the Big Bang but, rather, had a long prior history. Although the
details of this theory are far from understood, it is not
unreasonable to assume, since the Universe had to pass through the Big Bang
where densities and temperatures are set by the Planck scale, that
superstrings play a fundamental role. If this is the case, then one expects
the natural setting for cosmology to be, not four dimensions, but
the higher-dimensional spacetime of string theory. Ekpyrotic cosmology
introduced the idea that the Big Bang perhaps resulted from the
catastrophic collision of two brane solitons in this higher dimensional
space. Colliding branes and the associated bulk space geometry correspond to
vacuum solutions of the string theory equations of motion. These vacua may
solve equations which are valid to some finite order in a string expansion parameter,
such as five-brane/nine-brane collisions in heterotic $M$-theory~\cite{cosmoD,cosmoE} which are
valid to order $\kappa_{11}^{2}$, or they may be
exact conformal field theories, which solve the string equations to
all orders in the string parameter $\alpha'$. The second type of vacua were
emphasized in the Big Crunch/Big Bang~\cite{Khoury:2001bz} realizations of Ekpyrotic theory. Cyclic
models~\cite{cosmoF} are based on Big Crunch/Big Bang theories. All of these vacua
have in common the property that, prior to the Big Bang, the relevant region
of spacetime, our past, is contracting toward a singular brane collision.
In Big Crunch/Big Bang scenarios, the Universe then expands outward from this
singularity as the Big Bang. The singular point is called the ``bounce''.

Of particular importance in all cosmologies, and no less so in Ekpyrotic
theories, is the origin and momentum spectra of both scalar field and
gravitational quantum fluctuations. These are of the utmost importance since
they produce inhomogeneities in the cosmic microwave background (CMB) that have
already been observed. These observations are being increasingly refined and
offer an experimental window into whatever fundamental physics is responsible
for the present Universe. It was shown in~\cite{cosmoA,cosmoC} that nearly scale invariant
scalar perturbations are generated in the Big Crunch phase of Ekpyrotic
theories prior to the Big Bang. Indeed, a near scale invariant spectrum of
perturbations is observed in the CMB, but this is in the post Big Bang phase.
To make contact with these observations, one one must conclusively demonstrate
that the pre Big Bang fluctuation spectrum in Ekpyrotic theories
is propagated, nearly unchanged, through the Big Bang. This, despite the fact that
Big Crunch/Big Bang geometries are singular at the bounce. Although motivated
by string theory, almost all previous attempts to address this issue have been
carried out within the context of four-dimensional toy model geometries,
which do not solve the string equations of motion, even to lowest order~\cite{cosmoC,no,maybe}.
Recently, this problem was studied in a five-dimensional Milne background~\cite{Tolley:2002cv,
Tolley:2003nx}, and it
was emphasized that the five-dimensional structure is important near the Crunch.
However, in all cases the problem was studied
using the techniques of quantum field theory, with no string theory input.
With no further information at their disposal, the authors of these papers had
to proceed by matching the relevant pre Big Bang and post Big Bang wavefunctions
at the bounce, where the geometry is highly singular. The choice of boundary
conditions is, by itself, conjectural, being motivated by differing physical
arguments. This is made all the harder by the singular nature of the geometry
at the bounce. The result is that some authors claimed that the pre Big Bang
scale invariant fluctuations are radically altered when they pass through
the singularity~\cite{no} while others claim they are unaltered~\cite{cosmoC,Tolley:2003nx}. The ambiguous
nature of these attempts was demonstrated in~\cite{maybe}, who showed that, although
highly restricted, alternative boundary conditions are possible, leading to
the contradictory claims in the literature. How, then, can one resolve this
ambiguity?

It seems clear from the previous discussion that string theory should lead to
a unique solution of this problem. The reason for this is that string theory
vacua are, in general, globally defined. First of all, their geometric
manifolds include both the Big Crunch and Big Bang regions of cosmology, in
addition to other regions. Secondly, the wavefunctions of these vacua are
defined everywhere on the geometry. That is, knowing a wavefunction in the Big
Crunch region, for example, uniquely specifies the wavefunction in the
Big Bang regime. Clearly, this exactly specifies the boundary conditions for
the wavefunctions at the singularity, completely resolving the ambiguities
present in previous work. In this paper, we will show that this is indeed the
case, at least in the limit of zero string coupling. Working within the
framework of both supercritical \cite{Myers:fv,Antoniadis:1988vi} and critical type~II
superstrings\footnote{Note that ``type~II'' refers to a fermionic string with a chiral
GSO projection.}, we present a class of Big Crunch/Big Bang
cosmological vacua, called ``generalized'' Milne orbifolds \cite{Craps:2002ii}.
The geometry of these vacua includes both the past Big
Crunch and the future Big Bang regions. There are, in addition, four other
regions, often called ``whiskers'': an additional early time region, an additional
late time region and two intermediate regions with closed timelike curves. The
intermediate regions connect to the Big Crunch/Big Bang
regions at the bounce singularity. These vacua are exact superconformal field
theories and our results are valid to all
orders in the string worldsheet parameter $\alpha'$ \cite{Craps:2002ii}.
In this paper, we will present our discussion within the context of supercritical
string theory, since the corresponding spacetimes are manifestly homogeneous and
isotropic. However, we can show that the results are identical for
certain solutions of critical string theory. These ten-dimensional vacua contain, in
addition to the generalized Milne directions, two non-compact
spatial directions described by the
two-dimensional ``cigar'' conformal field theory \cite{Witten:1991yr},
which we take to be two of the three large
space directions. These vacua are not homogeneous and
isotropic, but approach these properties, as closely as one likes,
if a parameter is taken to be small. The Milne orbifold can be obtained as a specific limit
of these generalized Milne orbifolds. Working within this context, we find
that the wavefunctions are, indeed, globally defined. This follows from the
fact that the generalized Milne vacua all descend from a ``covering space'' by
the process of cosetting out a gauge action and orbifolding. The invariant globally
defined wavefunctions on the covering space then descend to globally defined
wavefunctions on the generalized Milne orbifolds. Therefore, to all orders in
$\alpha'$, the boundary condition ambiguities inherent in previous work have been
resolved.

What, then, are the results for the fluctuation spectrum? We find that the
fluctuation spectrum in the Big Crunch regime is, in general, altered by its
passage through the Big Crunch/Big Bang singularity. The change in the spectrum
can be calculated unambiguously at any time after the bounce. In the far future,
it can be expressed by an explicit momentum and time-dependent function
$\Delta(\vec{k},t)$,
which multiplies the early time pre Big Bang fluctuation spectrum. In the
Milne orbifold limit, we find that $\Delta\rightarrow 1$ and, hence, the
Big Crunch fluctuation spectrum is preserved as it passes through the
singularity. This proves the conjecture first introduced in~\cite{cosmoC} and shown
within a five-dimensional field theory context in~\cite{Tolley:2003nx}. If this result survives
corrections due to gravitational backreaction and stringy effects, which we will
comment on at the end of this Introduction, it means that the fluctuation
spectrum of Ekpyrotic cosmology may well be consistent with observation.
However, for generalized Milne orbifolds $\Delta$ is not unity and the
fluctuation spectrum changes as it passes through the bounce singularity. We
will show that this change is entirely due to the existence of
the whisker regions. Specifically, the quantum mechanics of these stringy
regions is inextricably linked to the quantum mechanics of the Big Crunch/Big
Bang geometry. The point of linkage is at the bounce singularity, where the
whiskers and the Big Crunch/Big Bang regions touch. The result is that
there is, in general, particle production in the future Big Bang region, even
though the vacuum in the Big Crunch was empty. This particle creation affects
all of the correlation functions in the future. In particular, it affects the
two-point correlation function from which $\Delta$ is calculated. We find that
$\Delta$ is an explicit function of momentum, with both time-independent and
time-dependent components. This calculation exposes a subtlety that arises
whenever the geometry has whiskers of the type discussed in this paper. That
is, that the early time in-vacuum is ambiguous. If we define this vacuum to be
such that an observer in the Big Crunch regime sees no particles, then we find
the associated in-vacuum in the early time whisker is not completely fixed. This
ambiguity in the in-vacuum can be parameterized by two constrained complex functions
$\gamma(\vec k)$ and $\tilde\gamma(\vec k)$. The parameter $\gamma(\vec k)$
explicitly enters the expression for $\Delta$. One natural choice turns out
to correspond to $\gamma(\vec k)=1, \tilde\gamma(\vec k)=0$ and leads to particle
creation and a
deviation of $\Delta$ from unity. However, as $\gamma\rightarrow 0$, the
particle production in the Big Bang region decreases to zero and $\Delta\rightarrow 1$. Therefore,
for a large choice of in-vacua, the change in the fluctuation spectrum by the
singularity is small. This might open the possibility of measuring string effects,
namely the existence of whisker regions connected to the Big
Crunch/Big Bang geometry, as small momentum and time-dependent deviations
from the scale invariance of the inhomogeneities in the CMB. We
hope to discuss this further elsewhere.

The linkage of the quantum mechanics of the Big
Crunch/Big Bang and whisker regions has a second, unanticipated effect.
As viewed in the Hilbert space of the complete quantum mechanics, the in-vacuum
is a pure state. However, we find that it is ``entangled'', that is, contains
correlations between the late time Big Bang and whisker regions. Therefore,
tracing its density matrix over the states of the unobserved whisker,
one obtains a non-trivial density matrix in the observable Big Bang region.
That is, an observer in the Big Bang region, with no access to information
about the whiskers, finds himself in a mixed state. This state has
``entanglement''entropy, which manifests itself explicitly in the expression
for $\Delta$. If the entropy were zero, then the expression for $\Delta$
should be compatible with a Bogolubov transformation linking the ``in'' and
``out'' states of the Big Crunch and Big Bang regions. However, non-vanishing
entanglement entropy will ruin this compatibility. We find that our explicit
expression for $\Delta$ indicates non-vanishing entanglement entropy, except
in the limit $\gamma\rightarrow 0$ where the whisker effects decouple. It
follows that, not only are the changes in the spectrum
calculable, but they explicitly exhibit entropy induced by the
existence of the stringy whisker regions. This is not dissimilar to recent
discussions of entangled states within the context of BTZ black holes
\cite{Maldacena:2001kr}, see also \cite{BTZ}.

We have presented the results of this paper mostly in the context
of Ekpyrotic Big Crunch/Big Bang transitions. Indeed, Ekpyrotic
theory has inspired much of the recent interest in Big Crunch/Big Bang
singularities in string theory \cite{Khoury:2001bz,Seiberg:2002hr}
and presents a context where it is particularly
important to know what happens to fluctuations at such a singularity.
However, Big Crunch/Big Bang transitions are of more general
interest and are especially appealing in string theory, where
observables are usually defined using simple asymptotic regions.
See, for instance, \cite{Gasperini:2002bn}.
Therefore, we believe that the results of this paper, such as
information loss due to the existence of  whisker regions, might be of interest in
more general cosmological scenarios.

Specifically, in this paper we do the following. Section 2 is devoted to a
discussion of a specific four-dimensional quantum field theory involving
gravity and three scalar fields. In subsection 2.1, we present a cosmological
background solution of this theory and explore various physical properties.
It is shown that there are two independent branches to
this solution, one representing a Big Crunch universe, evolving from the past
and ending in a singularity, and the other a Big Bang universe, beginning in a
singularity and evolving into the future. Scalar field quantum fluctuations, in
each of the Big Crunch and Big Bang regions, are discussed in subsection 2.2.
Both the in-vacuum and the out-vacuum are defined, and we explicitly compute
the scalar two-point correlation function with respect to each of
these vacua. In subsection 2.3, we relate both the classical and quantum
theories of the Big Crunch and Big Bang regions by connecting them at the
singularity. Within this context, we compute the two-point correlation
function with respect to the in-vacuum in both regions and compare them. We
find that the scalar fluctuation spectrum is potentially altered when it
passes from the past Big Crunch region through the bounce singularity. In the
far future, the change in the spectrum can be expressed by a momentum and
time-dependent function. We compute this function explicitly and show that it
depends on the Bogolubov coefficients relating the ``in'' and the ``out''
states of the Hilbert space. The meaning of these coefficients, and how one
should compute them, is discussed. Finally, in subsection 2.4, we show that
our four-dimensional quantum field theory is the low energy effective theory
for supercritical type II strings.

Section 3 is devoted to superstring cosmology within the context of type II
supercritical string theories. In subsection 3.1, we review the
``generalized'' Milne orbifold solutions of these theories first presented in
\cite{Craps:2002ii}.
These solutions are exact superconformal field theories. The various
regions of the associated spacetimes are discussed in detail, including the
additional whisker regions. It is shown that, at low energy,
these string vacua give rise to the Big Crunch/Big Bang theories introduced in
section 2. We study the scalar fluctuation spectrum in subsection 3.2. The
structure of generalized Milne orbifolds as the coset and orbifold of a
$PSL(2,\Rbar)$ covering space is reviewed and the relationship of these vacua
to the Milne orbifold is discussed. In subsection 3.2.1, a basis of
wavefunctions, each defined on every region of the spacetime, is presented and
studied. We quantize the scalar fluctuations in subsection 3.2.2. This is
accomplished by expanding the scalar fluctuations in this basis and
canonically quantizing the coefficients. We define two natural vacua,
the in-vacuum and the out-vacuum, and compute the Bogolubov coefficients
relating them \cite{Craps:2002ii}. Using these results, we compute the particle production in the
future regions and the scalar two-point correlation function with respect to
the in-vacuum. This result is valid in any region, allowing us to compare the
spectrum in the future Big Bang region with the early time spectrum
during the Big Crunch. We find that, generically, the spectrum is altered. In
the far future, we can express the change in the spectrum in terms of a
function $\Delta$, which is momentum and time-dependent. We compute this
function explicitly. In subsection 3.2.3, we show that there is, in fact, a
family of in-vacua, which we specify with two constrained complex functions
$\gamma(\vec k)$ and $\tilde\gamma(\vec k)$. We repeat the calculation of the Bogolubov coefficients, particle
production, scalar two-point correlation function and $\Delta$ in this context.
Again, in general, the fluctuation spectrum is changed as it passes through
the bounce singularity. However, in the limit that $\gamma \rightarrow 0$,
$\Delta$ approaches unity and the spectrum is conserved. In this limit,
$\Delta$ can be expressed in terms of the pure Big Crunch/Big Bang Bogolubov
coefficients. However, for any finite $\gamma$, this is no longer the case.
This is explained in subsection~3.3, where it is shown that the generalized
in-vacuum is an entangled state in the future. This implies that, from the point
of view of an observer in the Big Bang region, this vacuum has entanglement
entropy. This is equivalent to quantum mechanical ``information loss'' into
the whisker regions. It is this information loss that obstructs writing
$\Delta$ in terms of the pure Big Crunch/Big Bang Bogolubov coefficients. In
subsection 3.4, we discuss the limit of our superconformal vacua to the Milne
orbifold. It is shown that, in this limit, the factor $\Delta$ goes to unity
for any $\gamma$ in-vacuum. This proves that, in the Milne orbifold
cosmology, the fluctuation spectrum is unaltered by the bounce
singularity. In subsection~3.5, we include some comments on backreaction.
Finally, in Appendices A, B, and C, we outline the theory of
global wavefunctions, present the expressions for a basis of these
wavefunctions in all regions of the spacetime and discuss the Milne limit of
these wavefunctions respectively.

Before proceeding, we would like to make several important comments. First,
note that the generalized Milne orbifolds
have a smooth circle in the spatial Milne-direction.
However, it is not hard to show that all of the results of this paper will be
unchanged if we allow a further $\Zbar_{2}$ orbifolding in this direction. In
this case, this smooth circle becomes a finite interval, with new twisted
sector states appearing on each of the two boundaries. This new $\Zbar_{2}$
orbifolded vacuum is again an exact superconformal field theory, which is
closer in spirit to the notion of colliding branes.

As we have stated previously, our results are computed with respect to  exact
superconformal field theories, so $\alpha'$ corrections are under control.
However, it is well-known that these theories suffer from severe backreaction
problems, which could conceivably modify our results.
Indeed, string perturbation theory gives rise not only
to an expansion in $\alpha'$ (higher derivative corrections to the classical action),
but also to an expansion in the string coupling $g_s$
(quantum corrections). The results of \cite{Craps:2002ii, Elitzur:2002rt} on
global wavefunctions refer only to the first term in the $g_s$ expansion. It has,
in fact, been shown in \cite{Berkooz:2002je}, following \cite{Liu:2002ft}, that classical
string scattering amplitudes exhibit divergences associated with a large backreaction of
the spacetime geometry to small perturbations. For a non-perturbative manifestation of
gravitational instability, see \cite{Horowitz:2002mw}.
Large gravitational backreaction
(or the absence thereof) in this and related models \cite{Elitzur:2002rt, Cornalba:2002fi,
related}
was also studied in \cite{backreaction}.
The tree-level divergences of \cite{Berkooz:2002je, Liu:2002ft} indicate a breakdown of
string perturbation theory. That is, it is not consistent to
ignore $g_s$ corrections if $g_s$ is small but non-zero. For a non-technical discussion
of this, see \cite{Liu:2002yd}.
Therefore, there is at present no fully controlled computational framework, and it is
conceivable that our detailed results will receive significant corrections when
backreaction is taken into account. One may hope that at least certain important
qualitative features, such as the peculiar causal structure of the string solutions
and its effect on four-dimensional cosmology, will survive $g_s$ corrections. However, it
will take significant advances in string theory to either establish or refute this.

It has been suggested, see for instance \cite{Bachas:2002qt}, that string winding
modes that become light
near the bounce might play a role in resolving the singularity in
string theory. In the case of the Milne orbifold, the description
of these winding modes is somewhat complicated, because the size of the
Milne circle grows without bound away from the singularity.
See \cite{Berkooz:2003bs} for a very recent discussion and
\cite{Nekrasov:2002kf} for earlier work. However, in the generalized Milne
orbifold the radius of the Milne circle approaches a constant
asymptotically, and the vertex operators for the winding modes are
explicitly known \cite{Craps:2002ii}. In the present paper, we
will ignore winding modes, except to note that they become light
near the Big Crunch/Big Bang singularity and invalidate a four-dimensional description
there. They are heavy away from the bounce and, thus, do not appear in the
effective action describing our four-dimensional cosmology. However, we
should stress that the computations we do for the generalized Milne orbifold
can easily be extended to include winding modes
\cite{Craps:2002ii}. Of course, the most interesting effects of winding modes
should involve string interactions, and these have not yet been
computed for the generalized Milne orbifold.

In string theory, the observables are S-matrix elements. At zero string coupling, the
regime
in which we will be working, they are determined by a Bogolubov
matrix. Our strategy is to compute this matrix using the globally defined vertex operators
\cite{Craps:2002ii, Elitzur:2002rt}. More specifically, we calculate the Bogolubov
matrix to leading order in $\alpha'$ from the associated globally defined
wavefunctions. However, as was mentioned in \cite{Craps:2002ii},
the higher $\alpha'$ corrections modify this matrix in a very
trivial way, simply multiplying some of the entries by a phase.
Therefore, the particle creation rates we compute are exact to all
orders in $\alpha'$. In this paper, we will go a step beyond computing
S-matrix elements and compute correlation functions at a fixed finite time,
using the global wavefunctions we obtained from string theory.


For additional recent string theory work related to Big Crunch/Big Bang singularities,
see \cite{recent}.

\section{Four-Dimensional Cosmology}\label{sec:fourd}

In this section, we will explore the cosmological properties of a specific
four-dimensional quantum field theory coupled to gravity.
This theory consists of three scalar
fields, denoted by $\sigma_{T}$, $\sigma_{R}$ and $\phi$ respectively, coupled to
the usual Einstein gravity. The action for this theory is given by
\begin{equation}
S=\frac{1}{2\kappa_{4}^{2}}\int d^{4}x\sqrt{-g}{\cal{L}},
\label{eq:1}
\end{equation}
where
\begin{equation}
{\cal{L}}=R-\frac{1}{2}g^{\mu\nu}\partial_{\mu}\phi\partial_{\nu}\phi
-\frac{1}{2}g^{\mu\nu}\partial_{\mu}\sigma_{T}\partial_{\nu}\sigma_{T}
-\frac{1}{2}g^{\mu\nu}\partial_{\mu}\sigma_{R}\partial_{\nu}\sigma_{R}
-4Q^2e^{\phi}.
\label{eq:2}
\end{equation}
There are two dimensionful parameters in this action, Newton's constant
$\kappa_{4}^{2}/8\pi$ ($\kappa_4$ having the dimension of a length)
and a mass scale $Q$, which we will think of as small compared to the
momenta of interest.

\subsection{The Background Spacetime}\label{sub:background}

We will be interested in cosmological solutions of the equations of motion
of this theory that are spatially homogeneous and isotropic, that is, of the
Friedman-Robinson-Walker (FRW) type. We find the following general class of such
solutions. The scalar fields are independent of all spatial coordinates and
evolve as follows. The field $\sigma_{T}$ simply vanishes,
\begin{equation}
\sigma_{T}=0,
\label{eq:3}
\end{equation}
whereas
\begin{equation}
\sigma_{R}=\sqrt2\log|\tanh(Qt)|
\label{eq:4}
\end{equation}
and
\begin{equation}
\phi=2\phi_{0}-\log|\sinh(2Qt)|.
\label{eq:5}
\end{equation}
Here
$\phi_{0}$ is an arbitrary integration constant. These fields drive a
time-varying metric which is independent of all spatial coordinates and given
by
\begin{equation}
g_{\mu\nu}=a(t)^{2}\eta_{\mu\nu},
\label{eq:7}
\end{equation}
where
\begin{equation}
a^{2}=e^{-2\phi_{0}}|\sinh(2Qt)|
\label{eq:8}
\end{equation}
and $\eta_{\mu\nu}$ is the metric of flat Minkowski space. It follows from the
form of the metric that $t$ is conformal time.

What are the cosmological properties of this class of solutions? Perhaps the
most salient feature is that the conformal factor of the metric, $a(t)^{2}$,
is monotonically decreasing in the time interval $-\infty < t < 0$ and
monotonically increasing for $0<t< +\infty$. We will refer to the spacetime in the
negative time interval as Region I and the spacetime for the positive time
interval as Region II. Let us be more specific about the geometry of these
regions. To make contact with the conventional analysis of FRW cosmologies, we
will first compute the equation of state
\begin{equation}
w=\frac{P}{\rho},
\label{eq:9}
\end{equation}
where $P$ is the pressure and $\rho$ the energy density. These quantities are
defined by the Einstein equations
\begin{equation}
{\cal{H}}^{2}=\frac{\kappa_{4}^{2}}{3}a^{2}\rho
\label{eq:10}
\end{equation}
and
\begin{equation}
{\cal{H}}'=-\frac{\kappa_{4}^{2}}{6}a^{2}(\rho+3P),
\label{eq:11}
\end{equation}
where
\begin{equation}
{\cal{H}}=\frac{a'}{a}
\label{eq:12}
\end{equation}
and the prime denotes differentiation with respect to conformal time $t$.
Using~\eq{eq:8}, we find from~\eq{eq:10} and~\eq{eq:11} that
\begin{equation}
\rho= \frac{3Q^{2}}{\kappa_{4}^{2}}e^{2\phi_{0}}\frac{\cosh^{2}(2Qt)}{|\sinh^{3}
(2Qt)|}
\label{eq:13}
\end{equation}
and
\begin{equation}
P=\frac{Q^{2}}{\kappa_{4}^{2}}e^{2\phi_{0}}\left(\frac{4-\cosh^{2}(2Qt)}{|\sinh^{3}
(2Qt)|}\right)
\label{eq:14}
\end{equation}
respectively. It follows that
\begin{equation}
w=\frac{1}{3}\left(\frac{4-\cosh^{2}(2Qt)}{\cosh^{2}(2Qt)}\right).
\label{eq:15}
\end{equation}
The first thing to notice is that $w$ is not constant in time. In the far past
and future
\begin{equation}
w\longrightarrow -\frac{1}{3}, \qquad t\longrightarrow \pm\infty.
\label{eq:16}
\end{equation}
As the cosmology evolves, one finds that
\begin{equation}
w=0, \qquad t_{0}=\pm\frac{1}{2Q}\log(2+\sqrt{3})\Leftrightarrow \cosh(2Qt_{0})=2
\label{eq:17}
\end{equation}
and
\begin{equation}
w=\frac{1}{3}, \qquad
t_{1/3}=\pm\frac{1}{2Q}\log(1+\sqrt{2})\Leftrightarrow
|\sinh(2Qt_{1/3})|=1.
\label{eq:18}
\end{equation}
Finally, as one approaches the origin, either from negative time or positive
time,
\begin{equation}
w\longrightarrow 1-\frac{16Q^{2}}{3}t^{2}+\cdots, \qquad t\longrightarrow \pm 0.
\label{eq:19}
\end{equation}
It follows that
\begin{equation}
w\cong 1, \qquad |t|\ll \frac{\sqrt{3}}{4Q}.
\label{eq:20}
\end{equation}
What is the interpretation of this somewhat unusual FRW cosmology? To
elucidate this, note that in a scalar dominated phase the energy density and
pressure are given by
\begin{equation}
\rho={1\over2\kappa_4^2}\left(\sum_{i=1}^{3}\frac{{\psi'_{i}}^{2}}{2a^2}+V\right),
\qquad
P={1\over2\kappa_4^2}\left(\sum_{i=1}^{3}\frac{{\psi'_{i}}^{2}}{2a^2}-V\right),
\label{eq:21}
\end{equation}
where $\psi_{i}, i=1,2,3$ are the fields $\sigma_{T}$, $\sigma_{R}$ and $\phi$
respectively and $V$ is the total potential energy.
Using~\eq{eq:2},~\eq{eq:3},~\eq{eq:4} and~\eq{eq:5},
it is easy to show that $\rho$ and $P$ in~\eq{eq:21} are identical to
expressions~\eq{eq:13} and~\eq{eq:14}. It follows that in both Regions I and
II our cosmology is
scalar dominated with the equation of state $w$ changing as the scalar fields
evolve. In the conformal time coordinate $t$, only in the regions described
by~\eq{eq:16} and~\eq{eq:20} does $w$ behave approximately as a constant, $w=-1/3$
and $w=1$ respectively.
Note that, in the latter time region, the conformal factor $a(t)$
in~\eq{eq:8} becomes
\begin{equation}
a\cong |\frac{t}{e^{2\phi_{0}}/2Q}|^{q}, \qquad q=\frac{1}{2},
\label{eq:22}
\end{equation}
which is consistent with constant $w=1$.

A second important feature of FRW cosmologies is the Hubble parameter
\begin{equation}
H=\frac{a'}{a^{2}}
\label{eq:23}
\end{equation}
and its inverse, $R_{H}=|H|^{-1}$, the Hubble radius. Using~\eq{eq:8}, one
finds that
\begin{equation}
H=Qe^{\phi_{0}}\frac{\cosh(2Qt)}{\sinh(2Qt)|\sinh(2Qt)|^{\frac{1}{2}}}.
\label{eq:24}
\end{equation}
Again, note that $H$ and, hence, $R_{H}$ are not constants but, rather,
evolve with conformal time $t$. Let us focus on the behaviour of the Hubble
radius. In the far past and future
\begin{equation}
R_{H}\longrightarrow \infty, \qquad t\longrightarrow \pm \infty.
\label{eq:25}
\end{equation}
The Hubble radius then changes monotonically, taking the value, for example,
\begin{equation}
R_{H}=\frac{e^{-\phi_{0}}}{\sqrt2Q}, \qquad t_{1/3}=\pm\frac{1}{2Q}
\log(1+\sqrt{2})
\label{eq:26}
\end{equation}
before behaving as
\begin{equation}
R_{H}\longrightarrow {|2Qt|^{3/2}e^{-\phi_0}\over Q}(1-Q^{2}t^{2}+\cdots),
\qquad t\longrightarrow \pm0
\label{eq:27}
\end{equation}
as the time approaches the origin. Hence,
\begin{equation}
R_{H}\cong {|2Qt|^{3/2}e^{-\phi_0}\over Q}, \qquad |t|\ll\frac{1}{Q},
\label{eq:28}
\end{equation}
which vanishes at $t=0$. Note that this behaviour is consistent with the
monotonic evolution of the conformal factor $a^{2}$ and is valid in both Region
I and II. From \eq{eq:8} and \eq{eq:28} it is clear that the Hubble radius goes to zero
more quickly than the physical wavelength of an excitation, which is proportional to $a$.
Therefore, all modes are ``frozen'' outside of the horizon at the Big Crunch/Big Bang transition.
Similarly, we conclude from \eq{eq:8} and \eq{eq:24} that at very early and very late times, modes
are inside or outside the horizon depending on whether their comoving momentum $\vec k$ satisfies
$\vec k^2>Q^2$ or $\vec k^2<Q^2$, respectively. In the following, we will focus on fluctuations with
$\vec k^2\gg Q^2$. Thus, these modes start out inside the horizon at very early times and freeze as they
approach the Big Crunch. Similarly, in the Big Bang region of spacetime they start out frozen
near the Big Bang and enter the horizon at some later time.

A third important quantity to consider is the scalar curvature, $R$.
In conformal time, the scalar curvature in FRW spacetimes
is given by
\begin{equation}
R=\frac{6a''}{a^{3}}.
\label{eq:29}
\end{equation}
It then follows from~\eq{eq:8} that
\begin{equation}
R=3Q^{2}e^{2\phi_{0}}\frac{\sinh^{2}(2Qt)-1}{|\sinh^{3}(2Qt)|},
\label{eq:30}
\end{equation}
an expression that is valid in both Regions I and II. Key values of the scalar
curvature occur at precisely the same times, namely  $\pm\infty$, $t_{0}$,
$t_{1/3}$ and $\pm0$ defined in~\eq{eq:17} and~\eq{eq:18} respectively,
that indicated the behaviour of the equation of state
$w$. In the far past and future
\begin{equation}
R \longrightarrow 0, \qquad t\longrightarrow \pm\infty.
\label{eq:31}
\end{equation}
As $|t|$ decreases from infinity toward zero, $R$ has the following
properties. To begin with, $R$ is positive
and increasing until
$|t_{0}|=\frac{1}{2Q}\log(2+\sqrt{3})$, after which $R$ monotonically decreases.
The scalar curvature vanishes at
\begin{equation}
R=0, \qquad  t_{1/3}=\pm\frac{1}{2Q}\log(1+\sqrt{2}).
\label{eq:32}
\end{equation}
For smaller values of $|t|$, $R$ is negative. As one approaches the origin,
\begin{equation}
R\longrightarrow
-\frac{3}{4}e^{2\phi_{0}}\frac{1}{|t|^{3}}(1-6Q^{2}t^{2}+\cdots), \qquad
t\longrightarrow \pm0
\label{eq:33}
\end{equation}
which diverges as
\begin{equation}
R\cong -\frac{3}{4}e^{2\phi_{0}}\frac{1}{|t|^{3}}, \qquad |t|\ll
\frac{1}{\sqrt{6}Q}.
\label{eq:34}
\end{equation}

The above results allow us to give a concise description of our
cosmological solution. First consider Region I, corresponding to the negative
time interval $-\infty<t<0$. The associated geometry is that of a spatially
homogeneous and isotropic FRW spacetime. In the distant past, the manifold has
vanishing scalar curvature and a divergent Hubble radius. As time progresses,
the scalar curvature first grows positively, reaches a maximum and then begins to
decrease, vanishing at a finite time $t_{1/3}$. Henceforth,
the curvature is negative, diverging as $t^{-3}$ as $t$ approaches the origin
$t=0$. Throughout the entire time interval $-\infty<t<0$, the Hubble radius is
monotonically shrinking from infinity to zero. Both the vanishing of the
Hubble radius and, particularly, the divergence of the scalar curvature as
$t \rightarrow 0$, tells us that Region I terminates abruptly at $t=0$.
Region I, therefore, is a classic example of what is called a ``Big Crunch''
cosmology. Region II, on the other hand, corresponding to the positive time
interval $0<t<+\infty$, is the exact mirror image of Region I
in the time direction. That is, Region II is a spatially homogeneous and
isotropic FRW spacetime that starts abruptly at $t=0$ with vanishing Hubble
radius and negatively infinite scalar curvature and then expands outward. The
scalar curvature increases from negative to zero to positive, reaches a
maximum and then decreases to zero as $t \rightarrow +\infty$. During the
entire time interval $0<t<+\infty$, the Hubble radius monotonically increases
from zero to infinity. Region II, therefore, is a classic example of a ``Big
Bang'' cosmology. It is essential to note that in general relativity,
because of the curvature singularity at $t=0$, there is no relationship between
Region I and Region II, each representing an independent cosmology.

\subsection{Fluctuations}\label{sub:fluctuations}

We now turn to a discussion of quantum fluctuations of the scalar fields
on the Big Bang/Big Crunch geometries presented above. To do this, we must first
expand $\sigma_{T}$, $\sigma_{R}$ and $\phi$ around their classical values,
which we now denote as $\langle  \sigma_{T}\rangle $,
$\langle  \sigma_{R}\rangle $ and $\langle  \phi\rangle $, given
in~\eq{eq:3},~\eq{eq:4} and~\eq{eq:5} respectively. That is,
\begin{equation}
\sigma_{T}=\langle  \sigma_{T}\rangle +\delta\sigma_{T}, \qquad
\sigma_{R}=\langle  \sigma_{R}\rangle +\delta\sigma_{R}, \qquad
\phi=\langle  \phi\rangle +\delta\phi.
\label{eq:35}
\end{equation}
Inserting each of these fields into its equation of motion, using
expressions~\eq{eq:7} and~\eq{eq:8} for the background metric and assuming
the fluctuations are of the form
\begin{equation}
\delta\sigma_{T}=\delta T(t) e^{i{\vec{k}}\cdot{\vec{x}}}, \qquad
\delta\sigma_{R}=\delta R(t) e^{ i{\vec{k}}\cdot{\vec{x}}}, \qquad
\delta\phi=\delta \Phi(t) e^{ i{\vec{k}}\cdot{\vec{x}}},
\label{eq:36}
\end{equation}
we find that
the fluctuation $\delta T$ is a solution of
\begin{equation}
\delta T''+2Q\coth(2Qt)\delta T'+{\vec{k}}^{2}\delta T=0,
\label{eq:37}
\end{equation}
$\delta R$ solves the same equation
\begin{equation}
\delta R''+2Q\coth(2Qt)\delta R'+{\vec{k}}^{2}\delta R=0
\label{eq:38}
\end{equation}
whereas
\begin{equation}
\delta\Phi''+2Q\coth(2Qt)\delta\Phi'+({\vec{k}}^{2}+4Q^2)\delta\Phi, \qquad
\delta\Phi\ll 1.
\label{eq:39}
\end{equation}
Note that the $\delta \Phi$ fluctuations will satisfy the same equation as
$\delta T$ and $\delta R$ for momenta
\begin{equation}
{\vec{k}}^{2}\gg 4Q^2.
\label{eq:40}
\end{equation}
Henceforth, we will restrict our discussion to this momentum regime. Since, in
this case, all three fluctuations are specified by the same equation, we will
simply focus on one of them, which we choose to be $\delta T$.

Let us search for solutions of the $\delta T$ fluctuation
equation~\eq{eq:37}. To do this, we must first specify the region of spacetime
in which we want to work. Begin by considering Region I, with negative conformal
time in the interval $-\infty<t<0$. For very early times, \eq{eq:37}
simplifies to
\begin{equation}
\delta T''-2Q\delta T'+{\vec{k}}^{2}\delta T=0.
\label{eq:41}
\end{equation}
It is easy to see that this has plane wave solutions of the form
\begin{equation}
\delta T_{\vec{k}}^{\pm}= C_{\vec{k}}e^{Qt}e^{\mp iE_{\vec{k}}t},
\label{eq:42}
\end{equation}
where
\begin{equation}
E_{\vec{k}}=\sqrt{{\vec{k}}^{2}-Q^{2}}
\label{eq:43}
\end{equation}
is the energy associated with momentum $\vec{k}$ and $C_{\vec{k}}$ is
a normalization constant. Note that $E_{\vec{k}}$ is a positive real
number in the momentum regime~\eq{eq:40} in which we are working. The
normalization constant can be determined using the scalar product
\begin{equation}
(\phi_{1},\phi_{2})=-i\int_{\Sigma}{\phi_{1}{\stackrel{\leftrightarrow}
{\partial}}_\mu\phi_{2}^{*} \sqrt{g_{\Sigma}}d\Sigma^{\mu}}
\label{eq:44}
\end{equation}
where $\Sigma$ is a space-like three-surface, $\sqrt{g_{\Sigma}}$ is the
volume element on that surface,
\begin{equation}
d\Sigma^{\mu}={\delta^{\mu}}_{0}\sqrt{-g^{00}}d{\vec{x}}
\label{eq:45}
\end{equation}
and $\phi_{i}$, $i=1,2$ are any two scalar functions. Using the metric given
in~\eq{eq:7} and~\eq{eq:8}, expression~\eq{eq:42} for $\delta T$
and~\eq{eq:44}, we find that
\begin{equation}
(\delta T_ {\vec{k}}^{\pm}e^{i\vec k\cdot x}, \delta
T_{\vec{k'}}^{\pm}e^{i\vec k'\cdot \vec x} )=\pm(2\pi)^3\delta(\vec k-\vec k')
|C_{\vec{k}}|^{2}
e^{-2\phi_{0}}E_{\vec{k}}. \label{eq:46}
\end{equation}
Note that the Klein-Gordon norm obtained from~\eq{eq:44} can be either positive or negative
depending on the frequency of the scalar function.
Setting the right hand side of~\eq{eq:46} equal to $\pm(2\pi)^3\delta(\vec k-\vec k')$,
it follows that the normalization constant, up to a phase, is given by
\begin{equation}
C_{\vec{k}}=\frac{e^{\phi_{0}}}{\sqrt{E_{\vec{k}}}}.
\label{eq:47}
\end{equation}
Therefore, the normalized asymptotic plane wave solutions of~\eq{eq:41}
are given by
\begin{equation}
\delta
T_{\vec{k}}^{\pm}=\frac{e^{\phi_{0}}}{\sqrt{E_{\vec{k}}}}e^{Qt}
e^{\mp iE_{\vec{k}}t}, \qquad t\longrightarrow -\infty.
\label{eq:48}
\end{equation}
One can, in fact, solve the $\delta T$ fluctuation
equation~\eq{eq:37} for any value of $t$ in Region I. The result is found to
be
\begin{equation}
\delta
T_{\vec{k}}^{+}= \frac{4^je^{\phi_{0}}}{\sqrt{E_{\vec{k}}}}
(-z)^{j}F(-j,-j;-2j;\frac{1}{z}),
\label{eq:49}
\end{equation}
where
\begin{equation}
j=-\frac{1}{2} + i\frac{E_{\vec{k}}}{2Q}, \qquad z=-\sinh^{2}(Qt)
\label{eq:50}
\end{equation}
and $F(a,b;c;x)$ is the hypergeometric function $_{2}F_{1}$ (see Appendix~B for
its definition and some properties). In addition,
the hermitian conjugate
\begin{equation}
\delta T_{\vec{k}}^{-}=\delta T_{\vec{k}}^{+*}
\label{eq:51}
\end{equation}
is an independent solution of~\eq{eq:37}.
Using the facts that
\begin{equation}
z \longrightarrow -\frac{e^{-2Qt}}{ 4}, \qquad
F(-j,-j;-2j;\frac{1}{z})\longrightarrow 1\ \ {\rm as}\ \  t\longrightarrow -\infty,
\label{eq:52}
\end{equation}
we see that~\eq{eq:49} and~\eq{eq:51} approach the plane waves solutions
\eq{eq:48} in the far past. One can show that~\eq{eq:49} and~\eq{eq:51}
diverge logarithmically at $t=0$.

Combining these results with the first
expression in~\eq{eq:36}, we see that any fluctuation $\delta\sigma_{T}$ in
Region I can be written as
\begin{equation}
{\delta\sigma_{T}}^{I}=\int\frac{d^3k}{(2\pi)^{3/2}}
({a_{\vec{k}}}^{I}\delta T_{\vec{k}}^{+}(t)
e^{i\vec{k}\cdot\vec{x}}+{a_{\vec{k}}}^{I*}\delta
T_{\vec{k}}^{-}(t)e^{-i\vec{k}\cdot\vec{x}}),
\label{eq:53}
\end{equation}
where ${a_{\vec{k}}}^{I}$ are arbitrary complex coefficients.
Note that the first function in this expansion, $\delta
T_{\vec{k}}^{+}(t)e^{i\vec{k}\cdot\vec{x}}$, corresponds to a pure positive
frequency plane wave in the far past. Similarly, in this limit, $\delta
T_{\vec{k}}^{-}(t)e^{-i \vec{k}\cdot\vec{x}}$ becomes a negative frequency plane
wave.

Thus far, our discussion of the fluctuations ${\delta \sigma_{T}}^{I}$ has been
strictly classical. However, the theory can be easily quantized by demanding
that ${\delta \sigma_{T}}^{I}$ and, hence, the coefficients
${a_{\vec{k}}}^{I}$ be operators in a Hilbert space. For simplicity,
we will continue to denote these operators by ${\delta
\sigma_{T}}^{I}$ and ${a_{\vec{k}}}^{I}$, suppressing the usual ``hat''
notation. The quantization will be
canonical if we assume that
\begin{equation}
[{a_{\vec{k}}}^{I}, {a_{\vec{k'}}}^{\dagger I}]=\delta^{3}(\vec{k}-\vec{k'}),
\qquad [{a_{\vec{k}}}^{I}, {a_{\vec{k'}}}^{I}]=[{a_{\vec{k}}}^{\dagger I},
{a_{\vec{k'}}}^{\dagger I}]=0.
\label{eq:55}
\end{equation}
The vacuum state of the quantum theory is then defined as the normalized state
$|0\rangle _{in}$ satisfying
\begin{equation}
{a_{\vec{k}}}^{I}|0\rangle _{in}=0
\label{eq:56}
\end{equation}
for all momenta $\vec{k}$. There are many objects that can now be discussed in
this context. In this paper, we will focus primarily on the two-point
correlation function
\be\label{eq:56bis}
_{in}\langle 0|{\delta \sigma_{T}}^{I}(t,\vec{x})
{\delta \sigma_{T}}^{I}(t,\vec{x}+\vec r)|0\rangle _{in}.
\ee
Using~\eq{eq:53},~\eq{eq:55} and~\eq{eq:56}, we find that this function is
independent of $\vec{x}$ and given by
\beqa
_{in}\langle 0|{\delta \sigma_{T}}^{I}(t,\vec{x})
{\delta \sigma_{T}}^{I}(t,\vec{x}+\vec r)|0\rangle _{in}&=&\int\frac{d^3k}{(2\pi)^{3}}
|\delta T_{\vec{k}}^{+}|^{2}e^{-i\vec k\cdot \vec r}\cr
&=&{1\over 2\pi^2}\int dk |{\vec{k}}|^2 |\delta T_{\vec{k}}^{+}|^{2} {\sin(|{\vec{k}}||{\vec{r}}|)\over
|{\vec{k}}||{\vec{r}}|}.
\label{eq:57}
\eeqa
In the following, we will always set $\vec r=0$ and consider
\begin{equation}
_{in}\langle 0|\delta{\sigma_{T}}^{I}(t,\vec{x})^{2}|0\rangle_{in}=\int{\frac{d^{3}k}{(2\pi)^{3}}|
\delta{T_{\vec{k}}}^{+}|^{2}},
\label{eq:57A}
\end{equation}
since this is sufficient for determining the fluctuation spectrum.
The result for general $\vec r$ can be obtained
by multiplying the integrand by $\sin(|\vec k||\vec r|)/|\vec k||\vec r|$.

Equation~\eq{eq:57A} can be computed for any time $t$ in Region I using the
expression given in~\eq{eq:49}. However, for our purposes, it is most
illuminating to evaluate it in the distant past. Inserting expression~\eq{eq:48},
we find that the correlation function~\eq{eq:57A} becomes
\begin{equation}
_{in}\langle 0|{\delta\sigma_{T}}^{I}(t,\vec x)^{2}|0\rangle _{in}=f(t)\int\frac{d^3k}{(2\pi)^{3}}
\frac{1}{2E_{\vec{k}}}
\label{eq:58}
\end{equation}
where
\begin{equation}
f(t)=2e^{2\phi_{0}}e^{2Qt}.
\label{eq:59}
\end{equation}
For ${\vec{k}}^{2}\gg Q^{2}$, the momentum regime~\eq{eq:40} in which we
are working, expression~\eq{eq:58} becomes, to next to leading order,
\begin{equation}
_{in}\langle 0|{\delta\sigma_{T}}^{I}(t,\vec x)^{2}|0\rangle _{in}=f(t)\int\frac{d^3k}{(2\pi)^{3}}
\frac{1}{|\vec{k}|}\left(\frac{1}{2}+\frac{Q^{2}/2}{2|\vec{k}|^{2}}\right).
\label{eq:60}
\end{equation}
The momentum dependence of the first term of the integrand is simply that of zero-point fluctuations in Minkowski space. On the other hand,
the second term corresponds precisely to a scale invariant fluctuation
spectrum. Note, however, that since $\vec{k}^{2} \gg Q^{2}$, the second term is subdominant to the Minkowski fluctuations. The
time-dependent factor $f(t)$ is not canonical. To
understand its origin, we note that the kinetic energy term for $\sigma_{T}$
in the Lagrangian~\eq{eq:2} is not canonically normalized in the
gravitational background given in~\eq{eq:7} and~\eq{eq:8}. This kinetic
energy term can be canonically normalized by defining a new scalar field
${\Sigma_{T}}^{I}$ as
\begin{equation}
{\Sigma_{T}}^{I}=e^{-\phi_{0}} |\sinh(2Qt)|^{1/2}\sigma_{T}.
\label{eq:61}
\end{equation}
Note that in the far past this expression becomes
\begin{equation}
{\Sigma_{T}}^{I}=f(t)^{-1/2}\sigma_{T}, \qquad t\longrightarrow -\infty,
\label{eq:62}
\end{equation}
where $f(t)$ is given in~\eq{eq:59}. It follows from this and~\eq{eq:60} that
at early times
\begin{equation}
_{in}\langle 0|{\delta
\Sigma_{T}}^{I}(t,\vec x)^{2}|0\rangle _{in}=\int\frac{d^3k}{(2\pi)^{3}}\frac{1}
{|\vec{k}|}\left(\frac{1}{2}+\frac{Q^{2}/2}{|\vec{k}|^{2}}\right).
\label{eq:63}
\end{equation}
We conclude that the factor $f(t)$
in the correlation function~\eq{eq:60} is simply a scale factor that can be
absorbed by a field redefinition or not, depending on taste. This concludes
our analysis of quantum fluctuations in Region I.

We now want to discuss the quantum fluctuations of $\delta \sigma_{T}$ in
Region II, that is, for conformal time in the positive interval $0<t<+\infty$.
Note that the fluctuation equation~\eq{eq:37} is independent of which region
is being considered. It follows that the analysis of quantum fluctuations in
Region II is essentially identical to that in Region I. For this reason, we
will simply present our results. To begin with, the fluctuations $\delta
T_{\vec{k}}^{+}$ and $\delta T_{\vec{k}}^{-}$ given in~\eq{eq:49}
and~\eq{eq:51} respectively remain a complete set of solutions of
equation~\eq{eq:37}. It follows that any quantum fluctuation in Region II can
be written as
\begin{equation}
{\delta \sigma_{T}}^{II}=\int\frac{d^3k}{(2\pi)^{3/2}}({a_{\vec{k}}}^{II}
\delta T_{\vec{k}}^{+}(t)e^{i \vec{k}\cdot\vec{x}}+{a_{\vec{k}}}^{II
\dagger} \delta T_{\vec{k}}^{-}(t)e^{-i \vec{k}\cdot\vec{x}}),
\label{eq:64}
\end{equation}
where ${a_{\vec{k}}}^{II}$ and ${a_{\vec{k}}}^{II \dagger}$ satisfy the
canonical commutation relations
\begin{equation}
[{a_{\vec{k}}}^{II},{a_{\vec{k'}}}^{II \dagger}]=\delta^{3}(\vec{k}-\vec{k'}),
\qquad [{a_{\vec{k}}}^{II}, {a_{\vec{k'}}}^{II}]=[{a_{\vec{k}}}^{II \dagger},
{a_{\vec{k'}}}^{II \dagger}]=0.
\label{eq:65}
\end{equation}
The vacuum state is then defined as the normalized state $|0\rangle _{out}$
satisfying
\begin{equation}
{a_{\vec{k}}}^{II}|0\rangle _{out}=0
\label{eq:66}
\end{equation}
for all momenta $\vec{k}$. Again, there are many objects that one may wish to
compute at this point. For example, in analogy with Region I, we find that in
the distant future
\begin{equation}
_{out}\langle 0|{\delta
\sigma_{T}}^{II}(t,\vec x)^{2}|0\rangle _{out}=g(t)\int\frac{d^3k}{(2\pi)^{3}}\frac{1}
{|\vec{k}|}\left(\frac{1}{2}+\frac{Q^{2}/2}{|\vec{k}|^{2}}\right),
\label{eq:67}
\end{equation}
where
\begin{equation}
g(t)=2e^{2\phi_{0}}e^{-2Qt}.
\label{eq:68}
\end{equation}
As discussed previously, the factor of g(t) arises from the
non-canonical normalization
of the $\sigma_{T}$ kinetic energy term in~\eq{eq:2} with respect to the geometric
background~\eq{eq:7} and~\eq{eq:8}. Proper normalization of this term can be
achieved by defining
\begin{equation}
{\Sigma_{T}}^{II}=e^{-\phi_{0}}|\sinh(2Qt)|^{1/2}{\sigma_{T}}^{II}
\label{eq:69}
\end{equation}
for any positive conformal time $t$. Note that in the far future this relation
becomes
\begin{equation}
{\Sigma_{T}}^{II}=g(t)^{-1/2}{\sigma_{T}}^{II}, \qquad t\longrightarrow +\infty,
\label{eq:70}
\end{equation}
where $g(t)$ is given in~\eq{eq:68}. In terms of this canonically normalized scalar
field, the fluctuation spectrum~\eq{eq:67} becomes
\begin{equation}
_{out}\langle 0|{\delta
\Sigma_{T}}^{II}(t,\vec x)^{2}|0\rangle _{out}=\int\frac{d^3k}{(2\pi)^{3}}\frac{1}{|\vec{k}|}
\left(\frac{1}{2}+\frac{Q^{2}/2}{|\vec{k}|^{2}}\right).
\label{eq:71}
\end{equation}
Therefore, the $g(t)$ factor in the
two-point correlation function~\eq{eq:67} is simply a scale factor. It
can be absorbed or not, depending on taste. This concludes our analysis of
quantum fluctuations in Region II.

As stated earlier, because of the curvature singularity at $t=0$, there is no
classical relationship between Region I and Region II. Each represents an
independent cosmology. The same is true for the quantum fluctuations that we
have just discussed. The creation operators
${a_{\vec{k}}}^{I \dagger}$ acting on the vacuum $|0\rangle _{in}$ create a
Hilbert space of states ${\cal{H}}^{I}$ representing the quantum theory in the
Big Crunch geometry of Region I. Similarly, the operators ${a_{\vec{k}}}^{II \dagger}$
acting on $|0\rangle _{out}$ create a Hilbert space
${\cal{H}}^{II}$ representing the quantum theory of the Big Bang geometry of
Region II. A priori, there is absolutely no relation between ${\cal{H}}^{I}$
and ${\cal{H}}^{II}$.

\subsection{Relating the Big Bang to the Big Crunch}\label{sub:BBBC}

It is the thesis of Ekpyrotic cosmology that the Universe did not begin at the
Big Bang. Rather, it  had a prior history, connected to our present geometry
via some catastrophic event. In the Big Crunch/Big Bang versions of Ekpyrotic
theory, this catastrophic event is a spacetime singularity at $t=0$. This
singularity is of the type found in the curvature scalar in Region I and Region II
as $t\rightarrow -0$ and $t\rightarrow +0$ respectively. We will, therefore,
within the context of the theory described by~\eq{eq:2}, construct a Big
Crunch/Big Bang scenario by connecting Region I and Region II classically at
the singular point $t=0$. Having done this, one must also specify a relation
between the quantum theories on these two regions. The most naive approach,
and the one we will adopt in this section, is to identify the two Hilbert
spaces. That is, assume that
\begin{equation}
{\cal{H}}^{I}={\cal{H}}^{II} \equiv {\cal{H}}.
\label{eq:72}
\end{equation}
One consequence of this is that the creation/annihilation operators
${a_{\vec{k}}}^{I}$,${a_{\vec{k}}}^{I
\dagger}$ and ${a_{\vec{k}}}^{II}$,${a_{\vec{k}}}^{II \dagger}$ all act on the
same Hilbert space ${\cal{H}}$. In ``normal'' quantum field theory, that is,
when there is no geometric singularity or event horizon separating the past
from the future, the ``out'' creation/annihilation operators are linearly
related to the ``in'' creation/annihilation operators via a so-called
Bogolubov transformation. We will assume that the same is true in our theory,
despite the existence of a curvature singularity at $t=0$. That is, we
postulate that
\begin{equation}
\left(\begin{array}{c} {a_{\vec{k}}}^{II \dagger} \\ {a_{-\vec{k}}}^{II} \end{array}
\right)
=\left(\begin{array}{cc} X^{*}(\vec k) & Y^{*}(\vec k) \\ Y(-\vec k) & X(-\vec k) \end{array}\right)
\left(\begin{array}{c} {a_{\vec{k}}}^{I \dagger} \\ {a_{-\vec{k}}}^{I} \end{array}
\right).
\label{eq:73}
\end{equation}
The complex Bogolubov coefficients $X$ and $Y$ are not completely
independent. They are constrained by the requirement that the
Region I and Region II creation/annihilation operators continue to
satisfy the canonical commutation relations~\eq{eq:55} and~\eq{eq:65}
respectively. It follows that
\begin{equation}
|X|^{2}-|Y|^{2}=1.
\label{eq:74}
\end{equation}
This is the only constraint on these coefficients. However, for simplicity, we
will further assume that
\be\label{assume}
X(\vec k)=X(-\vec k),\ \ Y(\vec k)=Y(-\vec k).
\ee
Relaxing these assumptions will not change any of our conclusions.

Having postulated the relations~\eq{eq:72} and~\eq{eq:73}, one can now
compute correlation functions that were not defined separately in Region I and
Region II. Specifically, we want to calculate the two-point function
\begin{equation}
_{in}\langle 0|{\delta \sigma_{T}}^{II}(t,\vec x)^{2}|0\rangle _{in},
\label{eq:75}
\end{equation}
where ${\delta \sigma_{T}}^{II}$ is the Region II field operator given in~\eq{eq:64},
whereas $|0\rangle _{in}$ is the Region I vacuum defined in~\eq{eq:56}. This is
easily accomplished using~\eq{eq:55},~\eq{eq:56},~\eq{eq:64},~\eq{eq:73}
and~\eq{eq:74}. The result is
$$
_{in}\langle 0|{\delta
\sigma_{T}}^{II}(t,\vec x)^{2}|0\rangle _{in}=
$$
\be\label{eq:76}
=\int\frac{d^3k}{(2\pi)^{3}}
\left((1+2|Y|^{2})|\delta T_{\vec{k}}^{+}|^{2}+
XY\delta
T_{\vec{k}}^{+2}+
X^{*}Y^{*}\delta T_{\vec{k}}^{-2}\right),
\ee
with the fluctuations $\delta T_{\vec{k}}^{+}$ and $\delta
T_{\vec{k}}^{-}$ given by~\eq{eq:49} and~\eq{eq:51} respectively.
Of particular physical importance is the
form of this correlation function in the distant future.
As $t\rightarrow +\infty$, expression~\eq{eq:76} becomes
\begin{equation}
_{in}\langle 0|{\delta
\sigma_{T}}^{II}(t,\vec x)^{2}|0\rangle _{in}=g(t)\int\frac{d^3k}{(2\pi)^{3}}\frac{1}
{|\vec{k}|}\left(\frac{1}{2}+\frac{Q^{2}/2}{2|\vec{k}|^{2}}\right)\Delta(\vec{k},t),
\label{eq:77}
\end{equation}
where the scale factor $g(t)$ is given in~\eq{eq:68} and
\begin{equation}
\Delta(\vec{k},t)=1+2|Y|^{2}+XYe^{-2iE_{\vec{k}}t}+X^{*}Y^{*}e^{2iE_{\vec{k}}t}.
\label{eq:78}
\end{equation}
Written in terms of the scalar field ${\Sigma_{T}}^{II}$ defined
in~\eq{eq:69}, and using~\eq{eq:70}, this simplifies to
\begin{equation}
_{in}\langle 0|{\delta\Sigma_{T}}^{II}(t,\vec x)^{2}|0\rangle _{in}=\int\frac{d^3k}{(2\pi)^{3}}
\frac{1}{|\vec{k}|}\left(\frac{1}{2}+\frac{Q^{2}/2}{2|\vec{k}|^{2}}\right)\Delta(\vec{k},t).
\label{eq:79}
\end{equation}

We want to compare these results to another correlation function, namely
\begin{equation}
_{in}\langle 0|{\delta \sigma_{T}}^{I}(t,\vec x)^{2}|0\rangle _{in}.
\label{eq:80}
\end{equation}
Note that, in this case, both the field operator ${\delta \sigma_{T}}^{I}$ and the
vacuum $|0\rangle _{in}$ are the Region I quantities defined in~\eq{eq:53}
and~\eq{eq:56} respectively. For arbitrary time $t$, this two-point function
was evaluated in~\eq{eq:57} and its functional form in the limit
$t\rightarrow -\infty$ presented in~\eq{eq:60}. Finally, in terms of the
scalar field ${\Sigma_{T}}^{I}$ defined in~\eq{eq:61}, and
using~\eq{eq:62}, the $t\rightarrow -\infty$ limit of this
correlation function becomes
\begin{equation}
_{in}\langle 0|{\delta \Sigma_{T}}^{I}(t,\vec x)^{2}|0\rangle _{in}=\int\frac{d^3k}{(2\pi)^{3}}
\frac{1}{|\vec{k}|}\left(\frac{1}{2}+\frac{Q^{2}/2}{2|\vec{k}|^{2}}\right).
\label{eq:81}
\end{equation}
Although these expressions were first derived strictly within the context of
Region I, they remain valid for the complete Hilbert space ${\cal{H}}$. First
note that, for arbitrary conformal time $t$, correlation functions~\eq{eq:76}
and~\eq{eq:57} are identical only if the complex Bogolubov coefficient $Y=0$.
This remains true when comparing the $t\rightarrow +\infty$ limit of
$_{in}\langle 0|{\delta \sigma_{T}}^{II}(t,\vec x)^{2}|0\rangle _{in}$ given in~\eq{eq:77} to the
$t\rightarrow -\infty$ limit of $_{in}\langle 0|{\delta
\sigma_{T}}^{I}(t,\vec x)^{2}|0\rangle _{in}$ presented in~\eq{eq:56}. The simplest
comparison can be made between the future correlation function~\eq{eq:79} and
the past correlation function~\eq{eq:81}, since the irrelevant scale factors
have been removed. As stated earlier, the form of the argument of
the momentum integral in~\eq{eq:81} is that of Minkowski fluctuations plus a subdominant scale invariant contribution in the far past.
However, the
$\Delta(\vec{k},t)$ factor in~\eq{eq:79} potentially modifies the
fluctuation spectrum in the far future. If coefficient $Y=0$, then it follows
from~\eq{eq:78} that $\Delta=1$ and one again obtains the spectrum \eq{eq:81}.
However, if $Y\neq0$, then $\Delta\neq1$ and the fluctuation
spectrum is modified. What are the physical consequences of this?

A central assertion of Ekpyrotic Big Crunch/Big Bang theories is that a
scale invariant fluctuation spectrum is generated in the Big Crunch
geometry prior to the singularity which is then transmitted, without
modification, to the Big Bang geometry after the singularity. It is these scale
invariant fluctuations that are assumed to account for the observed
fluctuations in the cosmic microwave background. But is this true? Or is the
fluctuation spectrum modified by the presence of the singularity? There has
been considerable controversy regarding this, with some authors concluding
that the spectrum is transmitted unchanged, some authors claiming it is
greatly modified and further literature showing that this question is
ambiguous as posed, requiring more physics input to uniquely resolve it. All
of this literature has attempted to confront this issue by imposing explicit boundary
conditions to match the incoming and outgoing wavefunctions at, or near,
the singularity.
However, attempting to study the vicinity of a singularity is difficult since
one expects short distance effects to greatly modify the geometry and the
physics. This is the source of the ambiguities. Our point of view is
different. We see that the question of the persistence of the
spectrum through the singularity is precisely expressed by whether or not the
Bogolubov coefficient $Y$ vanishes. If it vanishes, the
spectrum is transmitted through the singularity into the future. If it does
not vanish, the spectrum is modified after the Big Bang. We
conclude that to study this problem, one must, unambiguously, compute the
Bogolubov coefficient $Y$. But how? One could, of course, attempt to compute
$Y$ by matching the wavefunctions at the singularity. But, as we have said,
this process would be ambiguous. A far more concrete way would be to compute
$Y$ directly from string theory, where at least for certain types of singularities
global wavefunctions can be unambiguously defined.
Given such globally defined wavefunctions, one can perform calculations
in the asymptotic past and future, far away from the singularity at $t=0$.
This is the approach we will follow in the remainder of this paper.
However, to compute the Bogolubov coefficient $Y$ in this manner, it is necessary
to demonstrate that the field theory defined by~\eq{eq:2} is, in fact, the low
energy four-dimensional effective theory of some specific string theory. This
is indeed the case, as we will now show.

\subsection{Four-Dimensional Cosmology from String
Theory}\label{sub:four}

In ($d+1$)-dimensional spacetime, string theory gives rise to a classical
effective action of the form
\begin{equation}
S_{d+1}=\frac{1}{2\kappa_{d+1}^2}\int d^{d+1}x\sqrt{-g}e^{-2\Phi}(R_{d+1}+4g^{IJ}
\partial_{I}\Phi\partial_{J}\Phi-2\Lambda),
\label{eq:82}
\end{equation}
where $I,J=0,1,\dots,d$, $g_{IJ}$ and $\Phi$ are the ($d+1$)-dimensional string frame metric
and dilaton respectively and all other massless and massive string modes have
been set to zero. A positive tree level cosmological constant $\Lambda>0$
will arise, for example, in supercritical type II string theories in
\begin{equation}
d+1=26,42,58,\dots
\label{eq:83}
\end{equation}
dimensions \cite{Antoniadis:1988vi}.

Let us now assume the existence of solutions of the string
equations of motion with metrics of the form
\begin{equation}
g_{IJ}dx^{I}dx^{J}=g'_{\mu\nu}dx^{\mu}dx^{\nu}+e^{2\sigma_{T}'}\delta_{ab}
dx^{a}dx^{b}+e^{2\sigma_{R}'}(dx^{d})^{2},
\label{eq:84}
\end{equation}
where $\mu,\nu=0,1,2,3$, indices $a,b=4,\dots,d-1$ and $g_{\mu\nu}'$, $\sigma_{T}'$
and $\sigma_{R}'$ are all functions of the four-dimensional coordinates
$x^{\mu}$, $\mu=0,1,2,3$ only. In addition, we will take the dilaton $\Phi$ to
be a function only of these four-dimensional coordinates. The coordinates
$x^{a}$
parameterize a ($d-4$)-torus with a reference radius $r_{0}$. This radius will typically be chosen to be the string scale, $\sqrt{\alpha'}$, or perhaps a few orders of magnitude larger.
Similarly, the $d$-direction is compactified
on a circle or an interval, either having a reference radius which, for simplicity, we also choose to be $r_{0}$.  For length scales much larger than this radius,  all heavy modes
will decouple and we will arrive at a four-dimensional effective theory
describing the light modes. These light modes will necessarily include $g_{\mu\nu}'$, $\sigma_{T}'$, $\sigma_{R}'$ and $\Phi$. However, one should also make sure that there are no additional ``stringy'' light modes, such as winding modes around one of the internal circles.
For the moment, let us ignore this important subtlety, returning to it at the end of this section.
The associated action is most easily computed
from~\eq{eq:82} by Weyl rescaling the four-dimensional metric as
\begin{equation}
g_{\mu\nu}'=\Omega^{2} g_{\mu\nu},
\label{eq:85}
\end{equation}
where
\begin{equation}
\Omega^{2}=e^{2\Phi_{4}'}
\label{eq:86}
\end{equation}
and
\begin{equation}
\Phi_{4}'=\Phi-\frac{1}{2}(\sigma_{R}'+(d-4)\sigma_{T}').
\label{eq:87}
\end{equation}
Then, making the field redefinitions
\begin{equation}
\phi=2\Phi_{4}', \qquad \sigma_{T}=\sqrt{2(d-4)}\sigma_{T}', \qquad
\sigma_{R}=\sqrt{2}\sigma_{R}',
\label{eq:88}
\end{equation}
defining
\begin{equation}
\frac{1}{{\kappa_{4}}^{2}}=\frac{V_{T}V_{R}}{{\kappa_{d+1}}^{2}}
\label{eq:89}
\end{equation}
where $V_{T}$ and $V_{R}$ are the volumes of the $(d-4)$-torus and the
$d$-direction circle/interval respectively and dropping all higher derivative
terms, we find that the low energy action is given by
\begin{equation}
S=\frac{1}{2{\kappa_{4}}^{2}}\int d^{4}x\sqrt{-g}{\cal{L}},
\label{eq:90}
\end{equation}
with
\begin{equation}
{\cal{L}}=R-\frac{1}{2}g^{\mu\nu}\partial_{\mu}\phi\partial_{\nu}\phi
-\frac{1}{2}g^{\mu\nu}\partial_{\mu}\sigma_{T}\partial_{\nu}\sigma_{T}
-\frac{1}{2}g^{\mu\nu}\partial_{\mu}\sigma_{R}\partial_{\nu}\sigma_{R}
-2e^{\phi}\Lambda.
\label{eq:91}
\end{equation}
Note that this is exactly the four-dimensional theory introduced in~\eq{eq:1}
and~\eq{eq:2} if we set\footnote{In the simple supercritical
string models we are considering, the cosmological constant $\Lambda$ is of order the
string scale, which is at odds with our assumption \eq{eq:40} that
$Q$ is small compared to the momenta of interest. However, in the critical ``cigar geometry'' mentioned in the Introduction,
it is possible to choose $Q$ to be arbitrarily small. For this reason, we will simply ignore this issue in our analysis.}
\begin{equation}
Q=\sqrt{\frac{\Lambda}{2}}.
\label{eq:6}
\end{equation}
Recall, however, that this result is predicated on the assumption that there are no additional
stringy light modes. We must now check to see under what conditions this will be the case.

To analyze this, we must be more careful in our discussion of decoupling. It is clear
from~\eq{eq:84} that, in the string frame \eq{eq:82}, the physical radii of the $(d-4)$-torus
and the $d$-direction circle are
\begin{equation}
e^{\sigma_{T}'}r_{0}, \qquad e^{\sigma_{R}'}r_{0}
\label{eq:A}
\end{equation}
respectively. From~\eq{eq:3} and~\eq{eq:88} we see that, for the background geometries of
interest in this paper, $\sigma_{T}'$ vanishes and, hence, the physical $(d-4)$-torus
radius is time-independent and fixed at $r_{0}$. On the other hand, it follows from~\eq{eq:4}
and~\eq{eq:88} that the physical radius of the $d$-direction circle is time-dependent and given by
\begin{equation}
r_{R}(t)=|\tanh(Qt)|r_{0}.
\label{eq:B}
\end{equation}
For very early and late times, $r_{R}(t)$ approaches $r_{0}$. Note, however, that
\begin{equation}
r_{R}(t)\longrightarrow 0, \qquad t\longrightarrow \pm0.
\label{eq:C}
\end{equation}
Therefore, we must be concerned that stringy modes winding around the $d$-direction circle
will become very light as $t\rightarrow \pm0$, thus substantially changing the low energy
theory given in~\eq{eq:90} and~\eq{eq:91}. Can we estimate the time regime for which this
effective theory is no longer valid? To do this, note that the winding modes around the
$d$-direction circle typically have a mass
\begin{equation}
m_{winding}\cong \frac{r_{R}(t)}{\alpha'}.
\label{eqD}
\end{equation}
It then follows from~\eq{eq:B}, and the fact that $r_{0}$ is of the order of the string
scale or a few orders of magnitude larger, that the effective field theory~\eq{eq:90}
and~\eq{eq:91} will be valid for momenta $\vec k$ satisfying
\be\label{valid}
\vec k^2\ll{r_0^2\tanh^2(Qt)\over(\alpha')^2}.
\ee
Since we are interested in momenta satisfying%
\footnote{Massless modes with momenta below this scale
correspond to growing modes and were discussed in section~5 of
\cite{Craps:2002ii}. They lead to infrared divergences in string perturbation
theory which will be ignored in this paper but would be interesting
to understand better. We are not aware of any direct relation to the
divergences in string perturbation theory mentioned at the beginning of
section~\ref{sec:string}.}
\be\label{momQ}
Q^2< \vec k^2,
\ee
we have to make sure that there is a window of momenta satisfying both \eq{valid} and
\eq{momQ}. Since we imagine $Q$ to be much below the string scale, $\alpha'Q^2\ll1$,
and for $r_0^2\cong \alpha'$, such a window exists as long as
\begin{equation}
t^2\gg\alpha',
\label{eq:E}
\end{equation}
that is, as long as we stay more than a few string times away from the singularity.
Note that this  does not alter the discussion and
conclusions associated with the theory given in~\eq{eq:1} and~\eq{eq:2}. To begin with,
that theory is singular at the origin , so we could not really discuss the
$t\rightarrow \pm0$ regime, nor did we. Indeed, the stringy effects that descend
from~\eq{eq:82} in the time region~\eq{eq:E} could conceivably ``regulate'' this
singularity, making the theory well-defined at the origin. We will not explore
this possibility in this paper. Henceforth, we will simply recall that the low
energy theory~\eq{eq:1},~\eq{eq:2} or, equivalently, ~\eq{eq:90},~\eq{eq:91} is
not valid too close to the singularity.

Subject to this caveat, we conclude that the four-dimensional cosmology described in this
section descends from a class of string theories in $d+1$
dimensions. We will now examine these string theories and their cosmology in detail,
with the aim of eventually using them to compute the Bogolubov coefficients discussed above.


\section{String Theory Cosmology}\label{sec:string}

As we have just shown, the four-dimensional action \eq{eq:1},~\eq{eq:2} is
the low energy limit of the supercritical string theory \eq{eq:82}
dimensionally reduced to four dimensions on a $(d-4)$-torus times a circle.
The $(d+1)$-dimensional action~\eq{eq:82} is itself an effective theory, all
higher $\alpha'$, string loop and non-perturbative effects being ignored. These
 effects are important, and we will comment on them in the appropriate places
later in this paper. However, we will begin by studying the classical  string
action~\eq{eq:82} and, specifically, the $(d+1)$-dimensional cosmological
solutions of its equations of motion. One such solution, the ``periodically
identified generalized Milne solution'', or the ``generalized Milne orbifold''
for short, was first presented in \cite{Craps:2002ii}.
This turns out to be an exact solution of classical string theory, that is, it
does not receive $\alpha'$ corrections \cite{TseytlinJackJonesPanvel}. We will see that, upon dimensional
reduction to four dimensions, this solution includes the background spacetime
studied in subsection~\ref{sub:background}
This makes the generalized Milne orbifold an excellent context to try and answer the questions
raised in the previous section.

\subsection{The Background Spacetime}\label{sub:string:background}

\begin{figure}
\begin{center}
\epsfig{file=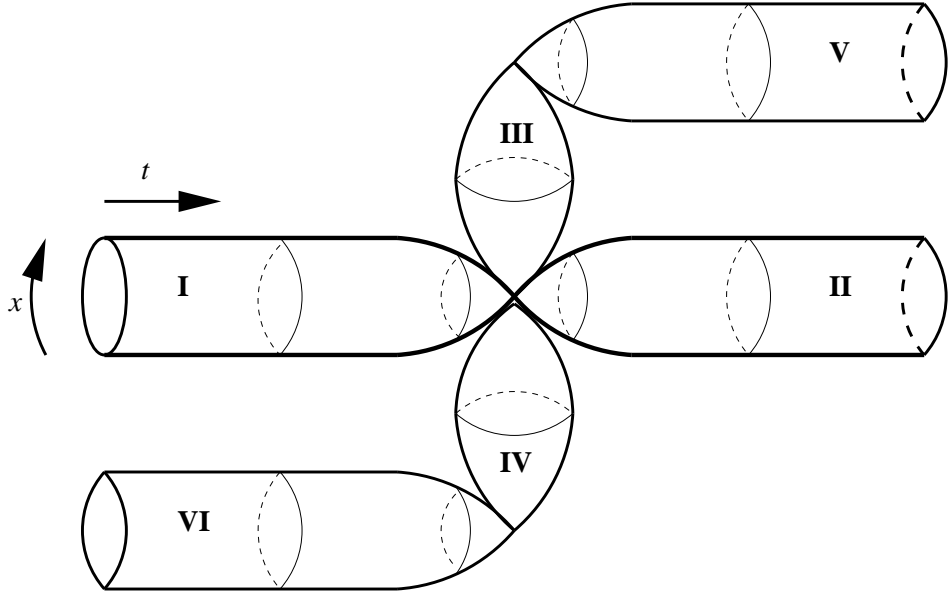, width=12.5cm}
\end{center}
\caption{The generalized Milne orbifold \cite{Craps:2002ii}.}\label{fig:CKR}
\end{figure}

Our starting point is the following solution to the
equations of motion associated with the classical string action \eq{eq:82}. The metric and dilaton are found to be
\beqa\label{backgroundIIA}
ds^2_{d+1}&=&{1\over Q^2}{du dv\over 1-uv}+
\sum_{I=1}^{d-1}(dx^I)^2,\cr
\Phi&=&-{1\over4}\log(1-uv)^2+\Phi_0.
\eeqa
Here
\begin{equation}
-\infty<u,v<\infty
\label{eq:F}
\end{equation}
are global two-dimensional coordinates with the identification
\be\label{uvid}
(u,v)\sim (u e^{-2\pi Q r_{0}}, v e^{2\pi Qr_{0}}).
\ee
The parameter $Q$ is related to the cosmological constant $\Lambda$
by expression \eq{eq:6}. The first term in the metric describes the
generalized Milne orbifold \cite{Craps:2002ii}. The remaining terms
describe  flat/torus-compactified $(d-1)$-dimensional space. The two-dimensional
generalized Milne directions exhibit an intricate causal structure that is schematically represented in \fig{fig:CKR}.

In Regions I and II of \fig{fig:CKR},
\begin{equation}
uv<0
\label{eq:G}
\end{equation}
and the identification \eq{uvid} is spacelike. These regions can also
be described by the coordinates $t,x$ defined by
\beqa\label{uvtx}
u=\sinh (Qt) e^{-Qx}, \qquad v=-\sinh (Qt)\, e^{Qx}
\eeqa
where
\be\label{xid}
x\sim x+2\pi r_{0}.
\ee
In these coordinates, the metric and dilaton solutions become
\beqa\label{backgroundIIAtx}
ds^2_{d+1}&=&-dt^2+\tanh^2(Qt)dx^2+\sum_{I=1}^{d-1}(dx^I)^2,\cr
\Phi&=&-\log\cosh(Qt)+\Phi_0.
\eeqa
For regions III and IV in \fig{fig:CKR},
\begin{equation}
0<uv<1.
\label{eq:H}
\end{equation}
The identification \eq{uvid} is now timelike. Thus, these regions contain closed timelike
curves. In regions V and VI, one has
\begin{equation}
1<uv
\label{eq:I}
\end{equation}
and the identification again becomes spacelike.

Region I describes a circle that starts out at some fixed radius
$r_{0}$, collapses to zero size and then, in Region II, expands again to radius $r_{0}$.
The Big Crunch/Big Bang singularity occurs when
\begin{equation}
uv=0.
\label{eq:J}
\end{equation}
At this singularity, regions I and II touch the
``whisker'' \cite{Elitzur:2002rt} regions IV and III which, in turn,
are connected to a second pair of asymptotic early and late time regions, VI and V respectively.
In addition, the separation boundary between regions VI and IV and between regions V and III, defined by
\begin{equation}
uv=1,
\label{eq:K}
\end{equation}
is singular. \fig{fig:CKR} is intended to give
a rough idea of the structure of the asymptotic regions, but
is less precise about the structure near the singularities.
For example, the metric in Region VI is actually
such that the radius of the circle is increasing
as one moves in towards the singularity. The string
coupling $g_s=\exp(\Phi)$ grows as well. In drawing  \fig{fig:CKR}, we have
implicitly performed a T-duality locally on that region,
to bring it to a form more like that of Region I.
Close to the singularity at $uv=0$, where regions I, II, III and IV meet,
the $u,v$ piece of \eq{backgroundIIA} reduces
to the Milne orbifold, a two-dimensional Minkowski space with the points
related by \eq{uvid} identified. See, for example, \cite{Cornalba:2002fi,
Nekrasov:2002kf, Berkooz:2002je,Berkooz:2003bs}
for recent discussions.
Deep into regions I and II, that is, as $uv\rightarrow -\infty$, the dilaton becomes
linear in time with the string coupling approaching zero.
At $uv=1$, the separation
between regions VI and IV and between regions V and III,
both the curvature and
the string coupling diverge.

A priori, one would
expect stringy, higher derivative corrections to \eq{eq:82}
to become important near the singularities at $uv=0$ and $uv=1$.
However, it was shown in  \cite{Craps:2002ii} that the background \eq{backgroundIIA}
defines an exact superconformal field theory with the correct central charge, $\hat c=10$ \footnote{The central charge deficit
of the $(u,v)$ directions is compensated by considering supercritical
string theory with a positive cosmological constant.}. That is, background~\eq{backgroundIIA} is an exact solution of the string equations of motion to all orders in $\alpha'$.

The coordinates $x^1,x^2,x^3$ range from $-\infty$ to $+\infty$.
The coordinates $x^a, i=4,\ldots,d-1$ are taken to be periodic, $x^a\sim
x^a+2\pi r_{0}$, so they describe a ($d-4$)-torus with constant radii $r_{0}$. These radii are chosen to be of the order of the string length or a few orders of magnitude larger.
As mentioned previously, the coordinate $x$ defined in~\eq{uvtx} is also compact although, in string frame, the
corresponding circle has a time-dependent radius $|\tanh(Qt)|r_{0}$.
The coordinate $t$ will be the time coordinate of an observer living in regions I and II.

We close this subsection by showing that in regions I and II of the $(d+1)$-dimensional cosmological solution given in~\eq{backgroundIIA}, there exists an effective
four-dimensional description of this background . This effective solution is valid everywhere except very close to the Big Crunch/Big Bang singularity, where
it is expected to break down due to additional
light stringy modes. It turns out that  the dangerous modes are
winding modes around the $x$ direction, which, as we have previously discussed,
indeed become light near $t=0$. With this caveat in mind, we proceed
to dimensionally reduce the metric and dilaton solutions in regions I and II
from $(d+1)$-dimensions to four dimensions by compactifying them on the $(d-4)$-torus times a circle. Recall that in regions I and II the metric and dilaton can be written as in~\eq{backgroundIIAtx}. Using the definitions in subsection~\ref{sub:four}, we find that, at low energy, \eq{backgroundIIAtx} corresponds to
\beqa\label{fourdeff}
\sigma_T&=&0,\cr
\sigma_R&=&\sqrt2\log|\tanh(Qt)|,\cr
\phi&=&2\Phi_0+\log 2-\log|\sinh(2Qt)|,\cr
g_{\mu\nu}&=&e^{-\phi}\eta_{\mu\nu},
\eeqa
which is exactly the four-dimensional background given in~\eq{eq:3},~\eq{eq:4},~\eq{eq:5} and~\eq{eq:7} of
subsection~\ref{sub:background} if we identify
\be\label{phizero}
\phi_0=\Phi_0+{1\over2}\log2.
\ee

It is crucial to note that \eq{backgroundIIAtx} only
describes regions I and II of the $(d+1)$-dimensional
spacetime \eq{backgroundIIA}. The full $(d+1)$-dimensional
string background contains two additional asymptotic regions
V and VI, as well as the intermediate regions III and IV.
The four-dimensional spacetime \eq{fourdeff} is a low energy description of
regions I and II only. However, as we have pointed out before, this
four-dimensional effective description breaks down near the Big
Crunch/Big Bang singularity, which is exactly where regions I
and II are connected to the other regions. To understand what
happens near the Big Crunch/Big Bang singularity, as well as in regions III, IV, V and VI, it is clearly necessary to use the full $(d+1)$-dimensional string theory background
\eq{backgroundIIA}. This is precisely what we will do throughout the remainder of this paper.

\subsection{Fluctuations}\label{sub:string:fluctuations}

In this subsection, we study fluctuations around the background
\eq{backgroundIIA}. The key ingredient in our discussion is that
string theory allows one to determine globally defined
wavefunctions, despite the singularities that prevent doing so in
general relativity \cite{Craps:2002ii,Elitzur:2002rt}.
The underlying reason is that the spacetime \eq{backgroundIIA}
corresponds to an orbifold of a coset conformal field theory.%
\footnote{For early work on applications of coset conformal field theories
to cosmology, see for example \cite{coset,Kounnas:1992wc}.}
There is a well-defined procedure to determine at least the
``untwisted'' globally defined vertex operators in such theories.
The zero mode parts of these vertex operators are the globally
defined wavefunctions we are interested in.

This procedure consists of two steps. In the first step, one
determines the vertex operators of the coset conformal field theory,
which in our case is $PSL(2,\Rbar)/U(1)$ at negative level.%
\footnote{$PSL(2,\Rbar)/U(1)$ at positive level corresponds to a
two-dimensional black hole geometry \cite{Witten:1991yr,Dijkgraaf:1991ba}.
$PSL(2,\Rbar)/U(1)$ at negative level can be
obtained from the black hole geometry by double Wick rotation \cite{Kounnas:1992wc}.
It is described by the $u,v$ piece of \eq{backgroundIIA}, without the
identification \eq{uvid}. In this spacetime, $u=v=0$ is a smooth
point (the Big Crunch/Big Bang singularity only arises after
the discrete identification \eq{uvid}), while there are
singularities at $uv=1$.}
One describes the coset conformal field theory as a gauged
Wess-Zumino-Witten (WZW) model. The vertex operators of the
ungauged $PSL(2,\Rbar)$ WZW model correspond to wavefunctions on
the smooth $PSL(2,\Rbar)$ group manifold. These can be found in
\cite{Vilenkin}.
The vertex operators of the $SL(2,\Rbar)/U(1)$ WZW model can then
be viewed as those vertex operators of the ungauged model that are
invariant under the $U(1)$ gauge group. See for example
\cite{Craps:2002ii, Elitzur:2002rt, Dijkgraaf:1991ba}.

The second step is the standard string theory orbifold procedure
\cite{orb}.
Roughly speaking, an orbifold is obtained from a
covering space, in our case the $PSL(2,\Rbar)/U(1)$ coset spacetime at negative
level,
by identifying points related by the action
of a discrete group. Here, the group is $\Zbar$ with the action
\eq{uvid}, which turns a line into a circle (the circles visible in
Fig.~\ref{fig:CKR}) and causes the Big Crunch/Big Bang singularity at
$uv=0$. The untwisted vertex operators%
\footnote{There are also twisted vertex operators, corresponding
to strings winding around the circles, see \cite{Craps:2002ii}.}
are those
vertex operators on the covering space that are invariant under
\eq{uvid}. From \eq{xid}, we see that this amounts to momentum
quantization in the $x$
direction. In this paper, we will only be
interested in zero momentum in the $x$ direction since the $x$
circle is part of the internal space. For the same reason, we will
not be interested in winding modes except to note that they cause
the effective four-dimensional description to break down near the
Big Crunch/Big Bang.

We would like to comment on the relationship of our theory to another model,
which has the same Big Crunch/Big Bang singularity. This is the Milne
orbifold $(\Rbar^{(1,1)}/\Zbar) \times \Rbar^8$, where $\Zbar$ is generated by
the boost transformation \eq{uvid} on two-dimensional Minkowski
space and $\Rbar^8$ denotes eight additional flat directions, some
of which may be compactified. The associated metric is
\be\label{Milne}
ds^2={1\over Q^2}dudv+\sum_{I=1}^8(dx^I)^2.
\ee
This spacetime is very similar to \eq{backgroundIIA} near the Big
Crunch/Big Bang singularity $u=v=0$, but differs significantly
from it away from this point, in particular in the structure of the
additional regions. The Milne orbifold consists of four cones
touching at $u=v=0$, that is, regions I, II, III and IV, but the
radius of the circle of the cones grows to infinite size as one
goes infinitely far away from the singularity. There are no
regions V and VI. They can be thought of as having been pushed
to infinity by focussing in on the manifold around $u=v=0$.

Wavefunctions on the Milne orbifold have been discussed from a
string theory point of view in \cite{Nekrasov:2002kf} and have
been used to compute string scattering amplitudes in
\cite{Berkooz:2002je}. There is no coset CFT involved in this
spacetime and the untwisted wavefunctions can be easily obtained from the
string orbifold procedure. That is, consider wavefunctions on Minkowski space
and demand that they be invariant under the action
\eq{uvid} of the orbifold group. Actually, such invariant
Minkowski space wavefunctions either grow or decay
in regions III and IV. Because these regions are non-compact
in the Milne orbifold, one, often implicitly, further restricts to
wavefunctions that decay in those regions. There
will be no such additional restriction in the spacetime
\eq{backgroundIIA}, since regions III and IV do not extend
to infinity there. As a consequence, more wavefunctions
are necessary in our theory than one finds in the Milne orbifold.

Global wavefunctions for the Milne orbifold with regions III and
IV omitted were presented in \cite{Tolley:2002cv} based on a
construction that does not refer to string theory, and applied in
\cite{Tolley:2003nx} to cosmology. The global wavefunctions agree
with the restriction to regions I and II of the stringy
wavefunctions discussed in the previous paragraph.
Also, the vacuum state implicitly defined by the wavefunctions of
\cite{Nekrasov:2002kf} and more explicitly used in \cite{Berkooz:2002je}
corresponds to the vacuum state defined
in \cite{Tolley:2002cv} upon deleting regions III and IV of
the string theory spacetime (or from the point of view of \cite{Tolley:2002cv},
upon adding those regions). It is of interest to see what happens to
the wavefunctions of the generalized Milne orbifold upon taking the limit
to the Milne orbifold. We analyze this in detail in subsection 2.4 and
Appendix~C.

The fact that the wavefunctions that descend from a covering conformal field
theory solution of string theory are globally defined on the associated
orbifold is of fundamental importance for the results of this paper. For that
reason, we outline in Appendix A, in more detail than discussed here, the procedure for
constructing these global wavefuntions.
In the remainder of this subsection, we will review and extend the
results of \cite{Craps:2002ii} on global wavefunctions and quantum
vacuum states in the spacetime \eq{backgroundIIA}.
The global
structure of the spacetime, in particular the presence of the
additional asymptotic regions, will be shown to have important
implications for the physics of the Big Crunch/Big Bang
transition.

\subsubsection{Global Wavefunctions on the Generalized Milne Orbifold}
\label{subsub:global}
We restrict our discussion to those fluctuations that are relevant for the
four-dimensional cosmology introduced in section~\ref{sec:fourd}. That is, the only
non-zero momentum components are in the three non-compact
space dimensions, and we ignore all winding and excited string
modes. In particular, momentum and winding in the $x$ direction
are set to zero,
\be\label{pw}
p=w=0
\ee
in the notation of \cite{Craps:2002ii}.
The justification for this is that all the modes we ignore have masses
on the order of the string scale, except very close to the Big Bang/Big Crunch singularity
where winding modes around the $x$ direction become light, as we
have previously mentioned.

In subsection~\ref{sub:fluctuations}, we solved  the wave equation
\eq{eq:37} for the fluctuations $\delta T$ defined in~\eq{eq:36}. Recall that from the point of view of the four-dimensional quantum field theory~\eq{eq:1} and~\eq{eq:2}, there were two classically independent regions, Region I and Region II.
In each of these regions, we found the same two independent solutions of~\eq{eq:37}
for a given momentum
$\vec k$. These solutions are ${\delta T_{\vec{k}}}^{+}$ in~\eq{eq:49} and
${\delta T_{\vec{k}}}^{-}(= {\delta T_{\vec{k}}}^{+*}) $.
Far from $t=0$, one solution, ${\delta T_{\vec{k}}}^{+}$, reduces to a positive
frequency wave, while its conjugate has negative frequency. However, as we have
shown in
subsection~\ref{sub:string:background}, this four-dimensional theory arises as
the low energy limit of the generalized Milne orbifold solution of the
$(d+1)$-dimensional classical string action~\eq{eq:82}. The full string theory
background \eq{backgroundIIA} has, in addition to regions I and II, the regions III,
IV,  V and VI discussed in detail above. Therefore, in string theory, one must solve the
$\delta T$ fluctuation equation in each of these six regions. Recall that
the generalized Milne orbifold arises from a covering $PSL(2,\Rbar)$ manifold by
constructing  the coset $PSL(2,\Rbar)/U(1)$ and then identifying points related by a
 $\Zbar$ group action. It follows from this structure that one can find solutions of
the $\delta T$ fluctuation equation on the generalized Milne orbifold by first finding
solutions of the fluctuation equation on $PSL(2,\Rbar)$ and then restricting to those
solutions that are invariant under the action of both $U(1)$ and $\Zbar$. Such
solutions, since they are globally defined on $PSL(2,\Rbar)$, remain globally
defined on the generalized Milne orbifold. That is, each such solution is defined
in each of the six regions I, II, III, IV, V and VI. As explained in
\cite{Craps:2002ii,Elitzur:2002rt}, there are
four independent wavefunctions of this type for each momentum $\vec k$,
 which we denote by
\begin{equation}
\label{Kuv}
\K_{++,\vec{k}}(uv),\ \K_{+-,\vec{k}}(uv),\ \K_{-+,\vec{k}}(uv)\ {\rm and}\
\K_{--,\vec{k}}(uv).
\ee
Note that, in addition to their $\vec{k}$ dependence, the argument of these
functions is the coordinate $uv$. More precisely, they depend not only on $uv$
but, also, on the region I, II, III, IV, V or VI. This latter dependence is suppressed
in \eq{Kuv}. In regions I and II,
it follows from~\eq{uvtx} that
\begin{equation}
uv=-\sinh^{2}(Qt).
\label{eq:n1}
\end{equation}
Similarly, in regions V and VI we can write
\begin{equation}
1-uv=-\sinh^{2}(Qt).
\label{eq:n1bis}
\end{equation}
Hence, in these regions the wavefunctions are dependent on $t$,
as we would expect them to be.
The four independent wavefunctions in~\eq{Kuv} can be written as
\begin{equation}
\K_{\pm\pm,\vec{k}}=\N K_{\pm\pm,\vec{k}},
\label{eq:n2}
\end{equation}
where $\N$ is a $\vec{k}$-dependent normalization constant given in~\eq{calN} and
the expressions for each of the four functions $K_{\pm\pm, \vec{k}}$ in the
six regions are given in Appendix B. That is, the independent wavefunctions
in~\eq{Kuv} are explicitly known.
The $\K_{\pm\pm, \vec{k}}$ are defined such
that they are orthonormal, with norm squared $\pm 1$, with respect
to the appropriate generalization of the Klein-Gordon inner product
\eq{eq:44}. The
generalization is that $\Sigma$ should be a global ``Cauchy''
surface, intersecting regions I and VI or II and V, which
turn out to be equivalent. The $K_{\pm\pm,\vec{k}}$ given in Appendix A have the same norm
up to a sign, so, by dividing out a common factor, we can obtain
the normalized $\K_{\pm\pm,\vec{k}}$. For example, $\K_{++,\vec{k}}$ coincides
with $\delta T_{\vec k}^+$ in Region I and vanishes in Region
VI, so it indeed has unit Klein-Gordon norm.

A key point of this paper is that the wavefunctions $\K_{\pm\pm,\vec{k}}$
are globally
defined on the generalized Milne orbifold. This follows from the fact that
they descend from globally defined functions on the $PSL(2,\Rbar)$ covering
space. The expressions for each of the four independent wavefunctions, in each
of the six regions, are presented in Appendix B. It is, however, useful at
this point to discuss several of these wavefunctions in more detail. First
consider $\K_{++, \vec{k}}$. In Region I, it is found to be
\begin{equation}
\K_{++, \vec{k}}=\delta {T_{\vec{k}}}^{+},
\label{eq:c1}
\end{equation}
where $ \delta {T_{\vec{k}}}^{+}$ is given in~\eq{eq:49}.
This function is purely positive
frequency in the asymptotic region $t\ll -1/Q$, while it diverges
logarithmically near the Big Crunch singularity, $t\to -0$. In Region II,
one has
\begin{equation}
\K_{++, \vec{k}}\propto (-z)^{-j-1}F(-j,-j;-2j;\frac{1}{z}),
\label{eq:c2}
\end{equation}
where $j$ and $z$ are defined in~\eq{eq:50}.
In this region, $\K_{++, \vec{k}}$ is
purely positive frequency asymptotically and diverges logarithmically
near the Big Bang as $t\to +0$. In the intermediate
Region III,
\begin{equation}
\K_{++, \vec{k}}\propto (uv)^{j}F(-j,-j;1;1-\frac{1}{uv}),
\label{eq:c3}
\end{equation}
which diverges logarithmically near the Big Crunch/Big Bang singularity at
$uv=0$, while it approaches a constant near the $uv=1$ singularity that separates
regions III and V. Similarly, in Region IV
\begin{equation}
\K_{++, \vec{k}}\propto (uv)^{-j-1}F(j+1,j+1;1;1-\frac{1}{uv}).
\label{eq:c4}
\end{equation}
This diverges logarithmically
near $uv=0$ and approaches a constant near $uv=1$. In
Region V, $\K_{++, \vec{k}}$ is
a mixture of positive and negative frequency components with equal amplitude in the
asymptotic region $uv\gg 1/Q^2$, while it approaches a constant near $uv=1$.
The exact expression in Region V is
\begin{equation}
\K_{++, \vec{k}}\propto F(-j,j+1;1;1-uv).
\label{eq:c5}
\end{equation}
Finally, and importantly, in Region VI
\begin{equation}
\K_{++, \vec{k}}=0.
\label{eq:c6}
\end{equation}

The global behavior of $\K_{+-, \vec{k}}$ can be obtained by replacing
$I\leftrightarrow VI$,
$II\leftrightarrow V$ and $uv\leftrightarrow 1-uv$ in the above expressions for
$\K_{++, \vec{k}}$. Of particular importance for us is the form of $\K_{+-,
\vec{k}}$ in regions I and VI. We find that
\begin{equation}
\K_{+-, \vec{k}}=0
\label{eq:c7}
\end{equation}
and
\begin{equation}
\K_{+-, \vec{k}}=\delta {T_{\vec{k}}}^{+},
\label{eq:c8}
\end{equation}
in regions I and VI respectively. Similarly, the
global structure of $\K_{--, \vec{k}}$ is obtained from that of $\K_{++,
\vec{k}}$ by replacing $I
\leftrightarrow II$, $V\leftrightarrow VI$ and ``positive frequency''
$\leftrightarrow$ ``negative frequency''. Specifically, we will use the fact
that
\begin{equation}
\K_{--, \vec{k}}=\delta {T_{\vec{k}}}^{+*}
\label{eq:c9}
\end{equation}
and
\begin{equation}
\K_{--, \vec{k}}=0
\label{eq:c10}
\end{equation}
in regions II and V respectively.
Finally, the behavior of $\K_{-+, \vec{k}}$ is
obtained from that of $\K_{--, \vec{k}}$ by replacing $I\leftrightarrow VI$,
$II\leftrightarrow V$ and $uv\leftrightarrow 1-uv$. In particular,
we find that
\begin{equation}
\K_{-+, \vec{k}}=0
\label{eq:c11}
\end{equation}
in Region II, whereas
\begin{equation}
\K_{-+, \vec{k}}=\delta {T_{\vec{k}}}^{+*}
\label{eq:c12}
\end{equation}
in Region V.


\subsubsection{Quantization}\label{subsub:quantization}

We would now like to quantize the scalar fluctuations
$\delta \sigma_{T}$ on the
generalized Milne orbifold. Generically, this can be done by
expanding
\beqa\label{expand}
&&\delta\sigma_T=\int{d^3k\over(2\pi)^{3/2}}\left(a_{1,\vec k}\K_{1,\vec k}(uv)
e^{i\vec k
\cdot\vec x}+a_{2,\vec k}\K_{2,\vec k}(uv)e^{i\vec k\cdot\vec x}\right.\cr
&&\ \ \ \ \ \ \ \ \ \ \ \ \ \ \  \ \ \ \ \ \  \left.
+a_{1,\vec k}^{\dagger}\K_{1,\vec k}^*(uv)e^{-i\vec k\cdot\vec x}
+a_{2,\vec k}^{\dagger}\K_{2,\vec k}^*(uv)e^{-i\vec k\cdot\vec x}\right),
\eeqa
where $\{\K_{1,\vec k}, \K_{2,\vec k}\}$ is some appropriate pair of
orthonormal wavefunctions.
As compared to \eq{eq:53}, we have twice as
many functions in the expansion \eq{expand}. The reason is that there are twice
as many independent global wavefunctions on the generalized Milne orbifold
as there are independent wavefunctions in the effective four-dimensional
theory.
We now impose the canonical commutation relations
\be\label{cancomm}
 [a_{i,\vec k},a_{j,\vec k'}^{\dagger}]=
\delta_{ij}\delta^3(\vec k-\vec k'), \qquad [a_{i,\vec k},a_{j,\vec k'}]=
[a_{i,\vec k}^{\dagger},a_{j,\vec k'}^{\dagger}]=0.
\ee
and define the vacuum state $|0\rangle$ by
\be\label{vacuumstate}
a_1|0\rangle=a_2|0\rangle=0.
\ee

To proceed, we now must ask the question: what is a natural vacuum state to choose, or equivalently,
what is a natural set of wavefunctions $\{\K_{1,\vec k}, \K_{2,\vec k}\}$? In
\cite{Craps:2002ii}, two natural vacua were defined and shown to be
inequivalent. The first vacuum state corresponds to the
choice
$\K_{1,\vec{k}}=\K_{++, \vec{k}},\ K_{2, \vec{k}}=\K_{+-, \vec{k}}$. The
associated fluctuation expansion is
\beqa\label{expandin}
&&\delta\sigma_T=\int{d^3k\over(2\pi)^{3/2}}\left(a^{I}_{\vec{k}}\K_{++,\vec
k}(uv)
e^{i\vec k
\cdot\vec x}+a^{VI}_{\vec k}\K_{+-,\vec k}(uv)e^{i\vec k\cdot\vec x}\right.\cr
&&\ \ \ \ \ \ \ \ \ \ \ \ \ \ \  \ \ \ \ \ \  \left.
+a^{I \dagger}_{\vec{k}}\K_{++,\vec k}^*(uv)e^{-i\vec k\cdot\vec x}
+a^{VI \dagger}_{\vec{k}}\K_{+-,\vec k}^{*}(uv)e^{-i\vec k\cdot\vec x}\right).
\eeqa
Recall that at early times, $\K_{++, \vec{k}}$ is purely positive frequency in
Region I and, from~\eq{eq:c6}, vanishes
in Region VI, and vice versa for the wavefunction
$\K_{+-, \vec{k}}$. It follows that
the vacuum state constructed from $a^{I}_{\vec{k}}$ and $a^{VI}_{\vec{k}}$
would indeed be called empty
by early time observers in regions I and VI. Therefore, we denote this state by
$|0\rangle_{in}$.
The second vacuum state, specified by $|0\rangle_{out}$, corresponds to the
choice
$\K_{1, \vec{k}}=\K^*_{--, \vec{k}}$ and $\K_{2, \vec{k}}=\K^*_{-+, \vec{k}}$
and the associated expansion
\beqa\label{expandout}
&&\delta\sigma_T=\int{d^3k\over(2\pi)^{3/2}}\left(a^{II}_{\vec k}
\K^*_{--,\vec k}(uv)
e^{i\vec k
\cdot\vec x}+a^{V}_{\vec k}\K^*_{-+,\vec k}(uv)e^{i\vec k\cdot\vec x}\right.\cr
&&\ \ \ \ \ \ \ \ \ \ \ \ \ \ \  \ \ \ \ \ \  \left.
+a^{II\dagger}_{\vec k}\K_{--,\vec k}(uv)e^{-i\vec k\cdot\vec x}
+a^{V\dagger}_{\vec k}\K_{+-,\vec k}(uv)e^{-i\vec k\cdot\vec x}\right).
\eeqa
Since, at late times, $\K^*_{--, \vec{k}}$ is purely positive frequency in Region
II and, from~\eq{eq:c10}, vanishes
in Region V, and vice versa for $\K_{-+, \vec{k}}$, this state would similarly be called empty
by late time observers in regions II and V.

The relation between $|0\rangle_{in}$ and $|0\rangle_{out}$ can be determined
from
\be\label{bogol}
\pmatrix{\K_{--,\vec{k}}\cr \K_{-+,\vec{k}}\cr \K^*_{--, \vec{k}}\cr
\K^*_{-+,\vec{k}}}=
\pmatrix{A&C&0&B\cr
         C&A&B&0\cr
         0&B^*&A^*&C^*\cr
         B^*&0&C^*&A^*}
\pmatrix{\K^*_{++, \vec{k}}\cr \K^*_{+-, \vec{k}}\cr \K_{++, \vec{k}}\cr
\K_{+-, \vec{k}}},
\ee
where the Klein-Gordon orthonormality of the wavefunctions implies that
$A$, $B$ and $C$ satisfy the conditions
\begin{equation}
|A|^2+|C|^2-|B|^2=1, \qquad AC^*+A^*C=0.
\label{consistency}
\end{equation}
For the special case \eq{pw} we are considering, one finds
\beqa\label{ABC}
A&=&-1,\cr
B=C&=&{1\over i\sinh\left({\pi E_{\vec{k}}\over 2Q}\right)}.
\eeqa
Note that $B$ is a function of $\vec{k}$ and that $B(\vec{k})=B(-\vec{k})$.
Then,~\eq{bogol} simplifies to
\begin{equation}
\pmatrix{\K_{--,\vec{k}}\cr \K_{-+, \vec{k}}\cr \K^*_{--, \vec{k}}\cr
\K^*_{-+, \vec{k}}}=
\pmatrix{-1&B&0&B\cr
         B&-1&B&0\cr
         0&-B&-1&-B\cr
         -B&0&-B&-1}
\pmatrix{\K^*_{++, \vec{k}}\cr \K^*_{+-, \vec{k}}\cr \K_{++, \vec{k}}\cr
\K_{+-, \vec{k}}},
\label{bogolsimp}
\end{equation}
which can be inverted to
\be\label{bogolsimpinv}
\pmatrix{\K^*_{++, \vec{k}}\cr \K^*_{+-, \vec{k}}\cr \K_{++, \vec{k}}\cr \K_{+-,
\vec{k}}}=
\pmatrix{-1&-B&0&-B\cr
         -B&-1&-B&0\cr
         0&B&-1&B\cr
         B&0&B&-1}
\pmatrix{\K_{--, \vec{k}}\cr \K_{-+, \vec{k}}\cr \K^*_{--, \vec{k}}\cr
\K^*_{-+, \vec{k}}}.
\ee
It is straightforward to check these relations using the formulas in
Appendix~B.
Inserting~\eq{bogolsimpinv} into~\eq{expandin}, we find the Bogolubov transformation
\be\label{bogolubov}
\pmatrix{a^{II\dagger}_{\vec k}\cr a^{V\dagger}_{\vec k} \cr a^{II}_{-\vec k}\cr
a^{V}_{-\vec k}}=
\pmatrix{-1&-B&0&B\cr
         -B&-1&B&0\cr
         0&-B&-1&B\cr
         -B&0&B&-1}
\pmatrix{a^{I\dagger}_{\vec k}\cr a^{VI\dagger}_{\vec k} \cr a^{I}_{-\vec k}\cr
a^{VI}_{-\vec k}}.
\ee

Note that the Bogolubov  transformation \eq{bogolubov}
mixes creation with annihilation operators. This
implies particle creation at late times \cite{Craps:2002ii}.
To see this, note that for each momentum $\vec{k}$
\be\label{particlesII}
_{in}\langle 0|a_{\vec{k}}^{II\dagger}a_{\vec{k}}^{II}|0\rangle_{in}=|B|^2={1\over\sinh^2\left(
{\pi E_{\vec k}\over2Q}\right)}
\ee
and
\be\label{particlesV}
_{in}\langle 0|a_{\vec{k}}^{V\dagger}a_{\vec{k}}^{V}|0\rangle_{in}=
|B|^2={1\over\sinh^2\left(
{\pi E_{\vec k}\over2Q}\right)}.
\ee
The physical interpretation is that, at late times, the vacuum
$|0\rangle_{in}$ has $1/\sinh^{2}(\pi E_{\vec{k}}/2Q)$ particles with momentum
$\vec{k}$ in Region I and the same number of particles with momentum $\vec{k}$
in Region V for each value of $\vec{k}$.
Note that the number of particles created goes to zero%
\footnote{At least for the modes \eq{pw} we are considering.
The generalization of \eq{particlesII} is \cite{Craps:2002ii}
\be |B|^2={\cosh^2(\pi
m)\over\sinh^2\left({\pi E\over 2Q}\right)} \ee
with \be m\equiv{1\over2}\left({n\over Qr_0}-{wr_0\over
Q\alpha'}\right),\ee
$n$ and $w$ being integers labelling momentum and winding in the
$x$ direction and $E$ being the energy.
}
in the limit that
\begin{equation}
E_{\vec{k}}/Q\longrightarrow \infty.
\label{eq:d1}
\end{equation}
We will use this result in subsection~\ref{subsec:milne}.

Let us now, within the context of the generalized Milne orbifold,
compute the analogue of expression \eq{eq:57}, that is, the two-point
correlation function
\begin{equation}
_{in}\langle 0|{\delta \sigma_{T}}(uv,\vec x){\delta \sigma_{T}}(uv,\vec x+\vec
r)
|0\rangle _{in}.
\label{eq:80bis}
\end{equation}
Since this function is correlated with respect to the in-vacuum, it is
simplest to expand the fluctuations as in~\eq{expandin}.
Using this and \eq{cancomm}, we find that
\beqa
&&_{in}\langle 0|{\delta \sigma_{T}}(uv,\vec x){\delta \sigma_{T}}(uv,\vec x+\vec r)
|0\rangle _{in}\cr
&&={1\over 2\pi^2}\int dk {|\vec{k}|}^2 \left(|\K_{++,\vec k}|^{2}+
|\K_{+-,\vec k}|^{2}\right)
\frac{\sin(|\vec{k}||\vec{r}|)}{|\vec{k}||\vec{r}|}.
\label{eq:80tris}
\eeqa
As in section 1, we will always set $\vec{r}=0$ and consider
\begin{equation}
_{in}\langle 0|{\delta \sigma_{T}}(uv,\vec x)^{2}|0\rangle _{in}
=\int{ \frac{d^{3}k}{(2\pi)^{3}} \left(|\K_{++,\vec k}|^{2}+
|\K_{+-,\vec k}|^{2}\right)}.
\label{eq:d2}
\end{equation}
Since our wavefunctions are globally defined,
\eq{eq:d2} is valid for $(u,v)$ in any of the six regions of the orbifold.
Of course, the explicit form of of $\K_{++,\vec k}$ and $\K_{+-,\vec k}$ change
from region to region and, hence, so will the expression for the correlation
function. Let us begin by calculating~\eq{eq:d2} in Region I. It follows
from~\eq{eq:c1} and~\eq{eq:c7} that
\be\label{eq:80trisone}
_{in}\langle 0|{\delta\sigma_{T}}(t,\vec x)^{2}|0\rangle _{in}=
\int \frac{d^{3}k}{(2\pi)^{3}}|\delta T_{\vec{k}}^{+}|^{2}.
\ee
The agreement with \eq{eq:57} illustrates that the in-vacuum
defined by \eq{expandin} is indeed the correct vacuum in Region~I. As
previously, for momenta $\vec{k}^{2}\gg Q^{2}$ in the distant past, the
two-point function in Region I becomes
\begin{equation}
_{in}\langle 0|{\delta\sigma_{T}}(t,\vec x)^{2}|0\rangle _{in}=\int\frac{d^3k}{(2\pi)^{3}}
\frac{1}{|\vec{k}|}\left(\frac{1}{2}+\frac{Q^{2}/2}{2|\vec{k}|^{2}}\right),
\label{eq:d3}
\end{equation}
where we have ignored a momentum-independent factor that can be
removed by a field redefinition. The same
expression for the correlation function and conclusions hold in Region VI.

Having established this, we now calculate the object of real interest, namely,
the two-point function~\eq{eq:d2} in Region II. Using~\eq{bogolsimpinv} and the
expressions for the wavefunctions given in the previous subsection, we find
that
\beqa\label{eq:80tristwo}
&&_{in}\langle 0|{\delta \sigma_{T}}(uv,\vec x)^{2}|0\rangle _{in} \cr
&&=\int {\frac{d^{3}k}{(2\pi)^{3}} \left((1+2|B|^{2})|\delta T_{\vec{k}}^{+}|^{2}
+|B(\vec k)|^{2}\delta T_{\vec{k}}^{+2}+|B(\vec k)|^{2}
\delta T_{\vec{k}}^{- 2} \right)}.
\eeqa
As $t\rightarrow +\infty$, this
expression becomes
\begin{equation}
_{in}\langle0|{\delta \sigma_{T}}(uv,\vec x)^{2}|0\rangle _{in}=
\int{\frac{d^{3}k}{(2\pi)^{3}}\frac{1}{|\vec{k}|}\left(\frac{1}{2}+
\frac{Q^{2}/2}{2|\vec{k}|^{2}}\right)\Delta(\vec{k},t)},
\label{eq:e1}
\end{equation}
where we have ignored a momentum-independent factor and
\begin{equation}
\Delta(\vec{k},t)=1+2|B|^{2}+|B|^{2}e^{-2iE_{\vec{k}}t}+|B|^{2}e^{2iE_{\vec{k}}t}.
\label{eq:e2}
\end{equation}
This should be compared with the field theory result given in~\eq{eq:78}.
Clearly, it is impossible to find
coefficients $X$ and $Y$ such that \eq{eq:78} reproduces \eq{eq:e2}.
This discrepancy has an interesting origin and interpretation, which will be
the subject of subsection~\ref{sub:string:informationloss}.

\subsubsection{A Family of Vacuum States}\label{subsub:family}

So far, we have considered the two vacuum states of
\cite{Craps:2002ii}. One of them is empty in each of the early time
regions I and VI, while the other is empty in the late time regions
II and V. Using these vacua, we found that there is
non-trivial particle creation in Region II, described by
\eq{particlesII}. These particles influence the two-point function
of the corresponding field, leading to the non-trivial factor of
$\Delta(\vec{k},t)$ in \eq{eq:e2}.
Upon reflection, however, one is led to wonder whether the choice of vacua of
\cite{Craps:2002ii} is really unique and, in particular, whether
there exists alternative vacuum states for which the early time
two-point function \eq{eq:80trisone} is less drastically altered
upon going through the Big Crunch/Big Bang transition.

Since we are interested in an observer who starts out in Region~I and ends
up in Region~II after the Big Crunch/Big Bang transition, we will continue
to impose the condition that any alternative in-vacuum state should be empty in
Region I. That is, the two-point function in Region~I should equal
\eq{eq:80trisone}. However, we will no longer impose a similar condition
in the other in-region, Region~VI, since our observer does not live there.
We now construct a family of such generalized in-vacua and compute, in each
case, the analogues of \eq{particlesII} and \eq{eq:e1}. These correspond
to quantities our observer could, in principle, measure in Region~II.

We continue to use the expansion \eq{expand} with
\be\label{K1++}
\K_{1, \vec{k}}=\K_{++,\vec{k}},
\ee
which has positive frequency in Region~I and vanishes in Region~VI. However,
we now allow a more general wavefunction for $\K_{2, \vec{k}}$, which vanishes in
Region~I but does not necessarily have positive frequency in Region VI.
The most general such wavefunction, for a given momentum, is an arbitrary
normalized linear combination of $\K_{+-}$ and $\K_{+-}^*$ specified by
\be\label{K2gen}
\K_{\gamma\tilde\gamma,\vec k}=
\gamma(\vec k)\K_{+-,\vec k}+\tilde\gamma(\vec k)
(\K_{+-,\vec k}+\K_{+-,\vec k}^*),
\ee
where $\gamma(\vec k)$ and $\tilde\gamma(\vec k)$ are complex numbers satisfying
\be\label{gammanorm}
|\gamma(\vec k)+\tilde\gamma(\vec k)|^2-|\tilde\gamma(\vec k)|^2=1.
\ee
For each
\be\label{gammatildegamma}
\gamma(\vec k)\neq0,
\ee
there exists at least one $\tilde\gamma$ such that \eq{gammanorm} is
satisfied. Note that $\gamma(\vec k)=1,\tilde\gamma(\vec k)=0$
corresponds to the natural in-vacuum
$|0\rangle_{in}$ defined using \eq{expandin}.
For simplicity, we will assume that
\be\label{assumegamma}
\gamma(\vec k)=\gamma(-\vec k),\ \ \tilde\gamma(\vec k)=\tilde\gamma(-\vec k),
\ee
although more general choices would not change our conclusions.
Henceforth, we will take
\begin{equation}
\K_{2,\vec{k}}=\K_{\gamma\tilde\gamma, \vec{k}}
\label{eq:e3}
\end{equation}
and use the expansion
\beqa
&&\delta\sigma_T=\int{d^3k\over(2\pi)^{3/2}}\left(a^{I}_{\vec{k}}\K_{++,\vec
k}(uv)
e^{i\vec k
\cdot\vec x}+a^{VI}_{\vec k}\K_{\gamma \tilde\gamma,\vec k}(uv)
e^{i\vec k\cdot\vec x}\right.\cr
&&\ \ \ \ \ \ \ \ \ \ \ \ \ \ \  \ \ \ \ \ \  \left.
+a^{I \dagger}_{\vec{k}}\K_{++,\vec k}^*(uv)e^{-i\vec k\cdot\vec x}
+a^{VI \dagger}_{\vec{k}}\K_{\gamma \tilde\gamma,\vec k}^{*}(uv)e^{-i\vec k\cdot\vec x}\right).
\label{eq:e4}
\eeqa

We can now repeat the computations of subsection~\ref{subsub:quantization}
for the new in-vacuum states $|0\rangle_{\gamma\tilde\gamma}$ defined by
\eq{expand}, \eq{vacuumstate}, \eq{K1++},\eq{K2gen} and~\eq{gammanorm}. We continue to
use the natural out-vacuum defined using \eq{expandout}. It follows from the
expressions for the wavefunctions given in Appendix B and~\eq{K2gen} that
\be\label{bogolsimpinvbis}
\pmatrix{\K^*_{++, \vec{k}}\cr \K^*_{\gamma\tilde\gamma, \vec{k}}\cr \K_{++,
\vec{k}}\cr
\K_{\gamma\tilde\gamma, \vec{k}}}=
\pmatrix{-1&-B&0&-B\cr
         -\gamma^* B&-(\gamma^*+\tilde\gamma^*)&-\gamma^* B&-\tilde\gamma^*\cr
         0&B&-1&B\cr
         \gamma B&-\tilde\gamma&\gamma B&-(\gamma+\tilde\gamma)}
\pmatrix{\K_{--, \vec{k}}\cr \K_{-+, \vec{k}}\cr \K^*_{--, \vec{k}}\cr
\K^*_{-+, \vec{k}}},
\ee
where $B$ is given in~\eq{ABC}. Inserting this into~\eq{eq:e4} and comparing
to~\eq{expandout}, we find the Bogolubov transformation
\be\label{bogolubovbis}
\pmatrix{a^{II\dagger}_{\vec k}\cr a^{V\dagger}_{\vec k} \cr a^{II}_{-\vec k}\cr
a^{V}_{-\vec k}}=
\pmatrix{-1&-\gamma^*B&0&\gamma B\cr
         -B&-(\gamma^*+\tilde\gamma^*)&B&-\tilde\gamma\cr
         0&-\gamma^*B&-1&\gamma B\cr
         -B&-\tilde\gamma^*&B&-(\gamma+\tilde\gamma)}
\pmatrix{a^{I\dagger}_{\vec k}\cr a^{VI\dagger}_{\vec k} \cr a^{I}_{-\vec k}\cr
a^{VI}_{-\vec k}}.
\ee
This can be inverted to give
\be\label{bogolubovtris}
\pmatrix{a^{I\dagger}_{\vec k}\cr a^{VI\dagger}_{\vec k} \cr a^{I}_{-\vec k}\cr
a^{VI}_{-\vec k}}=
\pmatrix{-1&B&0&- B\cr
        \gamma B&-(\gamma+\tilde\gamma)&-\gamma B&\tilde\gamma\cr
         0&B&-1&-B\cr
         \gamma^*B&\tilde\gamma^*&-\gamma^*B&-(\gamma^*+\tilde\gamma^*)}
\pmatrix{a^{II\dagger}_{\vec k}\cr a^{V\dagger}_{\vec k} \cr a^{II}_{-\vec k}\cr
a^{V}_{-\vec k}}.
\ee
Using~\eq{bogolubovbis}, we can now compute the occupation numbers
\be\label{particlesIIgen}
{}_{\gamma\tilde\gamma}\langle 0|a_{\vec{k}}^{II\dagger}a_{\vec{k}}^{II}|0\rangle_{\gamma\tilde\gamma}=
|\gamma B|^2=
{|\gamma(\vec k)|^2\over\sinh^2\left({\pi E_{\vec k}\over2Q}\right)}
\ee
in Region~II and
\be\label{particlesVgen}
{}_{\gamma\tilde\gamma}\langle 0|a_{\vec k}^{V\dagger}a_{\vec k}^V|0\rangle_{\gamma\tilde\gamma}=
|B|^2+|\tilde \gamma|^2
\ee
in Region~V.
For the reasons given at the beginning of this subsection, we are principally interested
in the number of particles in Region~II. We see from \eq{particlesIIgen} that
this number can be made arbitrarily small by choosing $\gamma(\vec k)$ to be close
to zero. Note, however, that we cannot set $\gamma(\vec{k})=0$, since this
would be inconsistent with the constraint~\eq{gammanorm}.

Next we compute
\beqa\label{eq:80tristwogen}
&&_{\gamma\tilde\gamma}\langle 0|{\delta \sigma_{T}}(uv,\vec x)^{2}
|0\rangle _{\gamma\tilde\gamma}\cr
&&=\int \frac{d^{3}k}{(2\pi)^{3}} \left(|\K_{++,\vec k}|^{2}+ |
\K_{\gamma,\tilde\gamma,\vec k}|^{2}\right).
\eeqa
In Region~I, $\K_{+-, \vec{k}}$ vanishes. It follows that $\K_{\gamma
\tilde\gamma, \vec{k}}$ also vanishes and, hence,
\eq{eq:80tristwogen} reduces to the familiar
result \eq{eq:80trisone}.
From \eq{KII} in Appendix~B, we see that $\K_{+-, \vec{k}}$ is purely imaginary
in Region~II.
In fact, this is also the case in the other regions that touch $uv=0$. Therefore,
$\K_{\gamma\tilde\gamma, \vec{k}}$ reduces to $\gamma\K_{+-, \vec{k}}$
in those regions. Using this,
\eq{eq:80tristwogen} takes the following form in Region~II,
\beqa\label{eq:80tristwogenbis}
&&_{\gamma\tilde\gamma}\langle 0|{\delta \sigma_{T}}(uv,\vec x)^{2}
|0\rangle _{\gamma\tilde\gamma}\cr
&&=\int \frac{d^{3}k}{(2\pi)^{3}} \left((1+2|\gamma B|^{2})\,
|\delta T_{\vec{k}}^{+}|^{2}
+|\gamma(\vec k)B(\vec k)|^2\,
\delta T_{\vec{k}}^{+2}+ |\gamma(\vec k)B(\vec k)|^2\,\delta
T_{\vec{k}}^{-2} \right).\cr &&
\eeqa
In the far future where $t\rightarrow+\infty$, and for $\vec k^2\gg Q^2$,
this expression becomes
\begin{equation}
_{\gamma\tilde\gamma}\langle 0|{\delta \sigma_{T}}(uv,\vec x)^{2}
|0\rangle _{\gamma\tilde\gamma}=\int
\frac{d^{3}k}{(2\pi)^{3}}\frac{1}{|\vec{k}|}
\left(\frac{1}{2}+\frac{Q^{2}/2}{2|\vec{k}|^{2}}\right)\Delta(\vec{k},t),
\label{eq:e5}
\end{equation}
where we have dropped a momentum-independent scale factor and
\begin{equation}
\Delta(\vec{k},t)=1+2|\gamma B|^{2}+|\gamma B|^{2}e^{-2iE_{\vec{k}}t}+
|\gamma B|^{2}e^{2iE_{\vec{k}}t}.
\label{eq:e6}
\end{equation}
In the limit $\gamma (\vec k)\to 0$, this approaches the early time result
\eq{eq:d3}. Therefore, for this limiting choice of in-vacuum, an observer in
Region~II
would say that the fluctuations in Region~I went through the Big Crunch/Big
Bang unchanged.
For this special limiting case, the result \eq{eq:e6} can be reformulated
in the framework of subsection~\ref{sub:BBBC}. It corresponds to the Bogolubov
coefficients $X$ and $Y$ with the properties
\be\label{XY}
|X|=1,\ \ Y=0\ \ \ \ \ {\rm in\ the\ limit}\ \gamma(\vec k)\to0.
\ee
However, this limiting case is the only one for which the framework of
subsection~\ref{sub:BBBC} can be made to reproduce \eq{eq:e6}. This
apparent discrepancy is the subject of the next subsection.


\subsection{Information Loss}\label{sub:string:informationloss}

It is useful to use the Bogolubov transformation~\eq{bogolubovtris} to
explicitly write the relation between the in-vacua
$|0\rangle_{\gamma\tilde\gamma}$ and the natural out-vacuum $|0\rangle_{out}$.
We find that (remember \eq{assumegamma})
\be\label{zerogammaout}
|0\rangle_{\gamma\tilde\gamma}=N_{\gamma\tilde\gamma}
\exp\left\{\int d^3k\left(\theta a^{II^\dagger}_{\vec{k}}a^{II^\dagger}_{-\vec{k}}
+\lambda a^{V\dagger}_{\vec{k}}a^{V\dagger}_{-\vec{k}}
+\mu a^{II\dagger}_{\vec{k}}a^{V\dagger}_{-\vec{k}}\right)\right\}\,
|0\rangle_{out},
\ee
where
\beqa\label{thetalambda}
\theta&=&-{\gamma^* B^2\over2(\gamma^*(1-B^2)+\tilde\gamma^*)},\cr
\lambda&=&{\tilde\gamma^*-\gamma^*B^2\over2(\gamma^*(1-B^2)+\tilde\gamma^*)},\cr
\mu&=&{\gamma^*B\over\gamma^*(1-B^2)+\tilde\gamma^*}.
\eeqa
To check this, use \eq{bogolubovtris} to verify that $|0\rangle_{\gamma\tilde\gamma}$
defined by \eq{zerogammaout} is indeed annihilated by $a^{I}_{\vec{k}}$ and
$a^{VI}_{\vec{k}}$ given that
$|0\rangle_{out}$ is annihilated by $a^{II}_{\vec{k}}$ and $a^{V}_{\vec{k}}$.
$N_{\gamma\tilde\gamma}$
is a normalization constant, which we will determine for a special case. It is useful
to keep in mind that $B$ given in~\eq{ABC} is purely imaginary.

For the natural in-vacuum $|0\rangle_{in}$ defined in \eq{expandin}, which corresponds to
choosing $\gamma(\vec k)=1,\tilde\gamma(\vec k)=0$, the coefficients
in~\eq{thetalambda} become
\beqa\label{thetaone}
\theta&=&-{B^2\over2(1-B^2)},\cr
\lambda&=&-{B^2\over2(1-B^2)},\cr
\mu&=&{B\over1-B^2}.
\eeqa
In this case, we find
\be\label{None}
N_{10}=\prod_{\vec k}\left(1-B^2(\vec k)\right)^{-1/2}\equiv
\exp\left\{ V\int{d^3k\over(2\pi)^3}\log\left(1-B^2(\vec k)\right)^{-1/2}\right\},
\ee
with $V$ the volume of space. For strictly infinity $V$, $N_{10}$ vanishes, so we will
assume $V$ to be large but finite instead.
Then \eq{None} can be derived as follows. Define the unitary matrix
\be\label{S}
S=\exp\left\{\int d^3k\left(B(a^{II}_{\vec{k}}a^{V}_{-\vec{k}}
+a^{II\dagger}_{\vec{k}}
a^{V\dagger}_{-\vec{k}}-a^{II}_{\vec{k}}a^{V\dagger}_{\vec{k}}
-a^{II\dagger}_{\vec{k}} a^{V}_{\vec{k}})\right)\right\}.
\ee
Using the formula
\be
e^{-A}Be^A=B+[B,A]+{1\over2!}[[B,A],A]+\cdots,
\ee
it is straightforward to verify that
\beqa\label{SS}
a^{I}_{\vec{k}}=-Sa^{II}_{\vec{k}}S^{-1}, \qquad
a^{VI}_{\vec{k}}=-Sa^{V}_{\vec{k}}S^{-1}.
\eeqa
This implies that, up to a phase
\be\label{Sinout}
|0\rangle_{in}=S|0\rangle_{out}.
\ee
The normalization factor \eq{None} can then be obtained by
determining the coefficient of $|0\rangle_{out}$ upon expanding the right
hand side of \eq{Sinout}.\footnote{In doing this, it is convenient to work with
rescaled oscillators $\alpha_{\vec k}\equiv\left((2\pi)^3/V\right)^{1/2}a_{\vec k}$,
which satisfy the canonical commutation relation $[\alpha_{\vec k},\alpha^\dagger_{\vec k}]=1$.}

As another special case, consider the limiting values $\gamma(\vec k)\to0$,
$\tilde\gamma(\vec k)\to\infty$. This was the limit that turned off particle creation
in Region~II, as we discussed at the end of subsection~\ref{subsub:family}.
This corresponds to
\beqa\label{thetazero}
\theta&\to&0,\cr
\lambda&\to&{1\over2},\cr
\mu&\to&0,
\eeqa
and
\be\label{zerogammaoutbis}
|0\rangle_{0\infty}\to N_{0\infty}
\exp\left\{\int d^3k\left({1\over2}(a^{V\dagger}_{\vec{k}}a^{V\dagger}_{-\vec{k}})\right)\right\}\,
|0\rangle_{\rm out}.
\ee
However, it turns out that in the limit \eq{thetazero} the normalization factor
$N_{0,\infty}$ vanishes, even for finite $V$.

We have not computed the normalization factor for generic values of $\gamma$ and
$\tilde\gamma$, although it could, in principle, be done in the same way. For instance,
if $\gamma$ and $\tilde\gamma$ are both real and positive, the generalization of \eq{S} is
\beqa\label{Sbis}
S&=&\exp\left\{\int d^3k\left(-\rho(a^{II}_{\vec{k}}a^{V}_{-\vec{k}}+
a^{II\dagger}_{\vec{k}}
a^{V\dagger}_{-\vec{k}}-a^{II}_{\vec{k}}a^{V\dagger}_{\vec{k}}
-a^{II\dagger}_{\vec{k}} a^{V}_{\vec{k}})\right.\right.\cr
&&\ \ \ \ \ \ \ \ \ \ \ \ \ \ \ \ \ \left.\left.+\mu(a^{V}_{\vec{k}}a^{V}_{-\vec{k}}
-a^{V\dagger}_{\vec{k}}a^{V\dagger}_{-\vec{k}})\right)\right\},
\eeqa
with
\beqa\label{sigmatau}
\cosh(2\mu)&=&\gamma+\tilde\gamma, \cr
\sinh(2\mu)&=&-\tilde\gamma, \cr
\rho {e^{2\mu}-1\over2\mu}&=&-\gamma B.
\eeqa
If $\gamma$ and $\tilde \gamma$ do not have the same phase, the exponent of $S$
also involves $a^{II}_{\vec{k}}a^{II}_{-\vec{k}}$ and
$a^{II\dagger}_{\vec k}a^{II\dagger}_{-\vec k}$.

The Hilbert space of states ${\cal H}$ we have been using can be viewed as
the tensor product
\be\label{tensor}
{\cal H}=\H^{II}\otimes\H^V
\ee
of a Hilbert space $\H^{II}$ associated with $a^{II}_{\vec{k}}$ and
$a^{II\dagger}_{\vec{k}}$ and a Hilbert space $\H^V$ associated with
$a^{V}_{\vec{k}}$
and $a^{V\dagger}_{\vec{k}}$. These two Hilbert spaces can be thought of as
associated with the out-regions~II and~V respectively. Note that $\H$ could equally well be
written as a tensor product
of Hilbert spaces associated with in-regions~I and~VI.
The out-vacuum can, accordingly, be expressed as
\be\label{outvac}
|0\rangle_{out}=|0\rangle_{II}\otimes|0\rangle_V.
\ee
It is clear
that each in-vacuum defined by \eq{zerogammaout} is a pure state in this
tensor product Hilbert space. With a pure state, one can
associate a trivial density matrix
\be\label{density}
\rho_{\gamma\tilde\gamma}=
|0\rangle_{\gamma\tilde\gamma}\,{}_{\gamma\tilde\gamma}\langle0|.
\ee
When one is only interested in computing correlation functions in
Region~II, which would be the case for a physicist living in
that region and trying to predict the result of experiments done
there, it is convenient (and equivalent) to use the density matrix obtained by
tracing \eq{density} over $\H^V$. This is given by
\be\label{densitytwo}
\rho^{II}_{\gamma\tilde\gamma}=\sum_i{}_V\langle i|\rho_{\gamma\tilde\gamma}
|i\rangle_V,
\ee
where $\{|i\rangle_{V}\}$ is an orthonormal basis of $\H^V$. Note that
$|0\rangle_{\gamma\tilde\gamma}\in{\cal
H}=\H^{II}\otimes\H^V$, whereas $|i\rangle_V\in\H^V$. Therefore,
\eq{densitytwo} is indeed a density matrix in Region~II, that is,
an element of $\H^{II*}\otimes\H^{II}$. Now, if $\mu$ in
\eq{zerogammaout} is nonzero (as it is, except in the limit
$\gamma\to0$), $|0\rangle_{\gamma\tilde\gamma}$ is an entangled state in
${\cal H}=\H^{II}\otimes\H^V$. When one traces the density matrix
of an entangled pure state over $\H^V$, one obtains a non-trivial
density matrix \eq{densitytwo} in Region~II. Therefore, an observer in
Region~II with no access to information about Region~V finds
himself in a mixed state, that is, a state with entropy. This entropy
is called ``entanglement entropy'' and reflects the ignorance
about the correlated Region~V. In the limit $\gamma\to0$, one again obtains a
trivial density matrix
\be\label{densityzero}
\rho^{II}_{0\infty}=|0\rangle_{II}\,{}_{II}\langle0|,
\ee
which is consistent with \eq{XY}.

This explains the discrepancy noted at the end of
subsection~\ref{subsub:family}. In subsection~\ref{sub:BBBC}, we
assumed unitary evolution from Region~I to Region~II, that is, a
pure state in Region~I evolving into a pure (squeezed) state in
Region~II. This assumption generically turns out to be
inconsistent with what we find in the full string theory
background, which includes additional regions. We find that a pure
in-vacuum state evolves to a pure state in the tensor product
Hilbert space $\H$, but that this state generically contains
correlations between regions~II and~V. As a consequence, in a
description where Region~V is ignored, such as the one
appropriate for an observer in Region~II, this state is a mixed
state. Only in the limit $\gamma\to0$ does the in-vacuum evolve to
a pure state in Region~II. As explained in
subsection~\ref{subsub:family}, the limit $\gamma\to0$ can,
strictly speaking, not be reached. The more accurate statement is that
the entropy of the state as described in Region~II can be made
arbitrarily small by taking $\gamma$ to be arbitrarily close to zero.

The above discussion is a generalization of thermofield dynamics as
recently applied to the eternal BTZ black hole, where the thermal
state of a scalar field outside the black hole is obtained from an entangled
state in a tensor product of two Hilbert spaces, the second Hilbert space
being associated with an additional asymptotic region behind the horizon
of the black hole \cite{Maldacena:2001kr}. Also see
\cite{BTZ} and references therein.


\subsection{The Milne Orbifold Limit}\label{subsec:milne}
In the limit that
\be\label{Milnelim}
uv\to0,
\ee
the metric in~\eq{backgroundIIA} reduces to the Milne orbifold metric \eq{Milne}
\be\label{Milnemetric}
ds^2={1\over Q^2}dudv+\sum^{8}_{I=1}(dx^I)^2
\ee
subject to the identification \eq{uvid},
\be\label{identorb}
(u,v)\sim(ue^{-2\pi QR},ve^{2\pi QR}).
\ee
In addition, it follows from~\eq{backgroundIIA} that, in this limit,
\begin{equation}
\Phi\longrightarrow \Phi_{0}.
\label{eq:h1}
\end{equation}
Note that the theory we are explicitly discussing in this
paper, \eq{backgroundIIA}, differs in the number of
compactified dimensions from the Milne orbifold
\eq{Milne}. Specifically, the generalized Milne orbifold has $(d+1)$-dimensions,
whereas the Milne space has ten dimensions. How, then, can we claim that
metric \eq{backgroundIIA} reduces near the singularity to \eq{Milne}? The
naive answer is that this limit does not involve the internal spatial
dimensions which, therefore, are irrelevant. A more precise way of
understanding the Milne limit is to recognize, as we mention in the
Introduction, that two flat spatial directions of the generalized Milne
orbifold can be replaced by a ``cigar'' geometry \cite{Witten:1991yr}
without changing any of the
conclusion of this paper. In this case, we have a critical, ten-dimensional
string theory where $Q$ is a free parameter. This cigar geometry has a
well-defined smooth limit to the Milne orbifold. Keeping this in mind, we will
ignore the number of compact dimensions in this subsection.

If in the limit \eq{Milnelim} we keep the ratio
\be\label{fixed}
\frac{{\vec{k}}^{2}uv}{Q^2} \quad {\rm fixed},
\ee
then the fluctuation equation \eq{eq:37} on the generalized Milne
orbifold reduces to
\be\label{eommilne}
\delta T''+{1\over t}\delta T'+\vec k^2\delta T=0.
\ee
Were we to consider fields with non-vanishing momentum along the $x$
circle as well, we would also be required to keep the product
\begin{equation}
Qr_{0} \quad {\rm fixed}.
\label{eq:g1}
\end{equation}
Expression~\eq{eommilne} is precisely the fluctuation equation for a
scalar mode in the Milne orbifold. The limit defined
by~\eq{Milnelim},~\eq{fixed} and~\eq{eq:g1}
takes the generalized Milne orbifold to the Milne orbifold.

It is instructive to see what our
global wavefunctions $\K_{\pm\pm, \vec{k}}$ reduce to in this limit. In
Appendix~C, we show that $\K_{++, \vec{k}}$ reduces to the standard
wavefunctions defining the adiabatic vacuum of the Milne orbifold,
that is, superpositions of positive frequency plane waves in Minkowski
space. They are Hankel functions in regions~I and~II and
modified Bessel functions that decay asymptotically in regions~III
and~IV, the expressions in
the four regions being related by analytic continuation. Near the Big
Crunch/Big Bang singularity, these wavefunctions diverge
logarithmically. On the other
hand, we find that
\begin{equation}
\K_{+-, \vec{k}}\longrightarrow 0
\label{eq:g2}
\end{equation}
everywhere in the Milne limit. More precisely, it
is always zero in Region I, is proportional to a finite Bessel function in
Region II and corresponds to an asymptotically growing modified Bessel function in
regions~III and~IV. Near the
Big Crunch/Big Bang singularity, $\K_{+-, \vec{k}}$ approachs constants in
all regions. However, its overall normalization factor vanishes, suppressing
the wavefunction completely. It then follows from~\eq{K2gen} that, in this limit,
\begin{equation}
\K_{\gamma\tilde \gamma, \vec{k}}\longrightarrow 0.
\label{eq:g3}
\end{equation}
Inserting these results into expression~\eq{eq:80tristwogen} for the scalar
two-point function, we find that, in the Milne limit, for any choice of
in-vacuum
\beqa\label{eq:f1}
_{\gamma\tilde\gamma}\langle 0|{\delta \sigma_{T}}(uv,\vec x)^{2}
|0\rangle _{\gamma\tilde\gamma}
=\int \frac{d^{3}k}{(2\pi)^{3}} |\K_{++,\vec k}|^{2},
\eeqa
which is valid in all regions. This is the standard result for
the two-point function in the adiabatic
vacuum of the Milne orbifold. It is useful to evaluate this correlation
function in regions I and II. It is straightforward to do this in Region I
using~\eq{eq:c1}. To evaluate~\eq{eq:f1} in Region II, first note
from~\eq{Milnelim}
and~\eq{fixed} that, in the Milne limit,
\begin{equation}
E_{\vec k}/Q=
\sqrt{\vec k^2-Q^2}/Q\to\infty
\label{eq:f2}
\end{equation}
and, hence, that the Bogolubov coefficient given in~\eq{ABC} satisfies
\begin{equation}
B\longrightarrow 0.
\label{eq:f3}
\end{equation}
One can then compute the correlation function in Region II using~\eq{eq:c9}
and the Bogolubov transformation~\eq{bogolsimpinvbis}. The result is that
in both regions I and II of the Milne orbifold,
\be\label{eq:f4}
_{in}\langle 0|{\delta\sigma_{T}}(t,\vec x)^{2}|0\rangle _{in}=
\int \frac{d^{3}k}{(2\pi)^{3}}|\delta T_{\vec{k}}^{+}|^{2}.
\end{equation}
It follows from our previous discussions that, in the distant past and distant
future, the spectrum is of the form of Minkowski fluctuations plus a subdominant scale invariant contribution and, therefore,
\begin{equation}
\Delta(\vec{k},t)=1.
\label{eq:g5}
\end{equation}
That is, in the Milne orbifold, the spectrum is unchanged by
passing through the Big Bang/Big Crunch singularity. Note that this quantum
process can be described by the pure Region I and Region II four-dimensional
effective theory with the Bogolubov coefficients
\begin{equation}
|X|=1, \qquad Y=0.
\label{eq:g6}
\end{equation}

Finally, we see that from~\eq{particlesII} and~\eq{eq:f3} that, in the Milne
limit,
\begin{equation}
_{in}\langle 0|a_{\vec{k}}^{II\dagger}a_{\vec{k}}^{II}|0\rangle_{in}
\longrightarrow 0.
\label{eq:g7}
\end{equation}
That is, particle creation is turned off.  We should mention that we have
assumed that $\gamma$ is kept fixed during the limiting proceedure.
It is clear that there would
be particle creation if, instead, one kept $|\gamma B|$ fixed during this limit.


\subsection{A Note on Backreaction}\label{sub:string:backreaction}

As we have emphasized at the end of the Introduction, the
string theory background \eq{backgroundIIA} has been argued to be
unstable to gravitational backreaction. Hence, reliable computations
should take this backreaction into
account. This seems to be out of reach at present and we will not attempt to resolve
the associated deep puzzles in string theory here. However, we would like to
offer a few remarks from the point of view taken in this paper.

First, it is worth translating the issue into the
four-dimensional language of section~\ref{sec:fourd}.
Looking at the energy density \eq{eq:13} of the
four-dimensional matter fields, one might be tempted to conclude
that quantum mechanical particles should not lead to large
backreaction. The reason is that the energy density of the classical matter
fields, since it is dominated by scalar kinetic energy,
scales like $|t|^{-3}$ near the Big Bang/Big Crunch singularity at $t=0$.
The wavefunction of a particle diverges at
most logarithmically near $t=0$, so, according to
\eq{eq:21}, its energy density cannot diverge more quickly than
that of the background matter fields. It would seem, therefore, that a
small fluctuation remains small compared to the background. So
where is the large backreaction?

The point is that the framework of section~\ref{sec:fourd} is
hard to use for computations near the bounce, since the geometry
is singular there. The right framework to use is
the one of section~\ref{sec:string}.
In these variables, there is not much energy in the classical
matter fields near the bounce, where spacetime is locally flat with
a slowly varying dilaton. The reflection of this in
section~\ref{sec:fourd} is that the sum of the energy densities in
the matter and gravitational fields diverges only like $1/|t|$, as
one can see from \eq{eq:13} and \eq{eq:30} (divided by $\kappa_4^2$).
This sum is dominated by the potential energy of the
four-dimensional dilaton $\phi$ which, in the string theory
framework, arises from the cosmological constant. Therefore, the criterion
for our computations to be unstable to backreaction is that
the energy density of small fluctuations grows faster than $1/|t|$
near the singularity. Unfortunately, this is the case for generic fluctuations, whose
wavefunctions diverge logarithmically at the bounce. Hence, we
recover the familiar large backreaction problem.

It is interesting to note that there is an important class of
fluctuations that do not give rise to large backreaction
\cite{Berkooz:2002je}. Of the modes considered in this paper, those
described by $\K_{+-, \vec{k}}$ are of this type, as can be seen using
Appendix~B. Their energy density actually vanishes at the Big Bang/Big
Crunch singularity.%
\footnote{This is not true for the $\K_{+-, \vec{k}}$ of the more generic
string modes considered in \cite{Craps:2002ii}. For example,
$\K_{+-, \vec{k}}$ does not
correspond to a chiral wavefunction near the singularity if it has
non-zero momentum along the $x$ circle.}

The issue of backreaction is crucial from various points of view,
string theoretic as well as cosmological. A source of criticism of the
simple orbifold-like cosmological singularities studied in the string
theory literature is that they are unstable against even a single particle
being added to the system before the singularity \cite{Horowitz:2002mw}. Such a
particle would change the conical singularity into a genuine curvature
singularity. In our model, this is indeed the case for particles with generic
wavefunctions, as we have just discussed. Now suppose that, however remote the
possibility, no such small fluctuation appeared and the Universe made
it through the Crunch singularity. Then, we have shown that the Big Crunch/Big Bang creates
a collection of particles with occupation numbers \eq{particlesIIgen} in
Region~II.
One can wonder how these particles backreact on the geometry. First, will the
energy density of these particles overwhelm the classical energy density very close
to the Big Crunch/Big Bang singularity? The answer is no, as can be seen from
\eq{eq:80tristwogen}. The
wavefunctions of the created particles are precisely proportional to $\K_{+-,
\vec{k}}$ near
the singularity, so their energy density vanishes there.%
\footnote{Again, this would be different for the modes we ignored in this paper. In order
to turn off their backreaction in Region~II, one would have to work in a $\gamma\to0$
limit.}
Second, it is tempting to speculate that, in some more realistic version of our model,
this particle creation might play a role as a reheating mechanism. In fact, reheating by
gravitational particle creation has been discussed before. See, for instance,
\cite{grav}.
\medskip

\section*{Acknowledgments}
We would like to thank C.~Armendariz-Picon, J.~Khoury and especially D.~Kutasov
for very useful discussions. We are also grateful to the organizers and participants of
the workshop ``Cosmological Perturbations on the Brane'', Cambridge, UK, for many
stimulating conversations. In addition, B.~C.\ would like to thank the High Energy
Theory Group at Penn for hospitality.
The work of B.~C.\ is supported in part
by DOE grant DE-FG02-90ER40560 and by NSF grant PHY-9901194. B.~A.~O.\ is
supported in part by DOE under contract No.\ DE-AC02-76-ER-03071.

\section*{Appendix A: Global Wavefunctions from String Theory}

In this Appendix, we will outline the procedure for constructing the globally
defined wavefunctions on both the Milne orbifold and the generalized Milne
orbifold discussed in this paper.

\paragraph{Milne Orbifold:}
We first review how globally defined wavefunctions are constructed on the
Milne orbifold $\Rbar^{1,1}/\Zbar\times\Rbar^8$, an orbifold of
Minkowski space.
\begin{enumerate}
\item
Solve the wave equation
\be\label{waveeq}
(\nabla^2+m^2)\phi=0
\ee
on the covering space $\Rbar^{1,9}$. A smooth basis of solutions is
given by the plane waves
\be\label{planewave}
\{e^{i\vec p\cdot\vec X}e^{i(p^+X^-+p^-X^+)}|2p^+p^--\vec p^2=m^2\}.
\ee
\item
Choose an alternative basis of solutions to \eq{waveeq}, consisting of
continuous
superpositions of the smooth solutions \eq{planewave}, such that the
orbifold generator $X^{\pm}\mapsto e^{\pm 2\pi}X^\pm$ is diagonal. Such a
basis is given by \cite{Nekrasov:2002kf} (see also \cite{Berkooz:2002je})
\be\label{Nekrwavef}
\phi_{l,\vec p}=e^{i\vec p\cdot\vec X}\int_{-\infty}^\infty dw
e^{i(p^+X^-e^{-w}+p^-X^+e^w)}e^{ilw}
\ee
with $l\in
\Rbar$. The orbifold generator acts by multiplication by
$e^{-2\pi il}$. The wavefunctions \eq{Nekrwavef} are generically not
smooth near the light-cone $X^+X^-=0$, which divides the covering
space in four regions. When studying string theory on Minkowski space,
one could in principle use these singular wavefunctions but, since
there exist smooth wavefunctions \eq{planewave} which can be written
as continuous superpositions of \eq{Nekrwavef}, it is preferable to
work with the latter.
\item
Restrict the wavefunctions \eq{Nekrwavef} to the orbifold invariant
ones, that is, to those with $l\in\Zbar$. Using coordinates $t,x$ defined by
$X^\pm=te^{\pm x}/\sqrt2$, in terms of which the orbifold generator
acts as $x\mapsto x+2\pi$, it is clear that $l$ is the momentum in the
$x$ direction. Therefore, the condition $l\in\Zbar$ is the usual momentum
quantization. Hence, we end up with wavefunctions \eq{Nekrwavef} on the
orbifold space. They are not smooth at the singularity $X^+X^-=0$, yet
they are globally defined. Note that, because of the quantization of
$l$, it is not possible to use a basis of smooth wavefunctions on the
orbifold space.
\item
Introduce twisted sectors (winding modes). We will not discuss those
in this paper; see \cite{Berkooz:2003bs} for a very recent discussion.
\end{enumerate}
\paragraph{Generalized Milne Orbifold:}
We now review the prescription of \cite{Elitzur:2002rt,Craps:2002ii} for
constructing global wavefunctions on the generalized Milne
orbifold, which in string theory corresponds to a $\Zbar$ orbifold of a
coset conformal field theory at negative level \cite{Craps:2002ii},
\be\label{genmil}
{PSL(2,\Rbar)_{k<0}\over U(1)}/\Zbar\  \times\Rbar^{d-1}.
\ee
\begin{enumerate}
\item
Start with wavefunctions on $AdS_3$, that is, the well-known
eigenfunctions of the Laplacian on $AdS_3$
which coincides with the $SL(2,\Rbar)$ group manifold. $SL(2,\Rbar)$ is the
group of $2\times2$ matrices $g$ with unit determinant,
\be
g=\pmatrix{a&b\cr c&d},\ \ \ ad-bc=1.
\ee
$PSL(2,\Rbar)$ is obtained from this by identifying $g$ with $-g$.
Wavefunctions on $PSL(2,\Rbar)$ should be consistent with this
identification. The wavefunctions on a group manifold coincide with
the matrix elements in the different representations in the group. Therefore,
we can write
\be\label{statesSL}
\phi_{j,\alpha,\beta}(g)=\langle j,\alpha|g|j,\beta\rangle,
\ee
where $j$ labels a $PSL(2,\Rbar)$ representation and $\alpha,\beta$
label states in the representation. In this paper, we will only
consider the principal continuous representations, with $j=-1/2+is$
and $s$ real. They correspond to taking $\vec k^2>Q^2$. A choice of basis of
wavefunctions now corresponds to a choice of states $\alpha,\beta$ in
the representation $j$. By choosing an appropriate basis, one can
choose a smooth basis of wavefunctions on $PSL(2,\Rbar)$. Note that
the group manifold $PSL(2,\Rbar)$ admits a $PSL(2,\Rbar)_L\times
PSL(2,\Rbar)_R$ symmetry, the two factors acting on $g$ by left and
right multiplication respectively. It is clear from \eq{statesSL}
that the action of $PSL(2,\Rbar)_L$ is determined by the state
$\langle j,\alpha|$, while the action of $PSL(2,\Rbar)_R$ is
determined by $|j,\beta\rangle$. The Laplacian on the $PSL(2,\Rbar)$
group manifold is the Casimir operator of $PSL(2,\Rbar)$ (acting either from
the left or from the right). Its eigenvalue is $j(j+1)$.
\item
Choose a basis of wavefunctions on $PSL(2,\Rbar)$ such that the
$U(1)_L\times U(1)_R$ subgroup, with both factors generated by
$\sigma_3$, is diagonalized. This amounts to choosing the basis
vectors of the representation $j$ such that $\sigma_3$ is diagonal. It
turns out \cite{Vilenkin} that, for the representations we are
considering, there are two states for each $\sigma_3$ eigenvalue
$m$.%
\footnote{To be precise,
$e^{\alpha\sigma_3}|j,m,\pm\rangle=e^{2im\alpha}|j,m,\pm\rangle$.}
Thus, we find four independent wavefunctions for each $j,m,\bar
m$
\be\label{KSL}
K_{\pm\pm}(j,m,\bar m;g)=\langle j,m,\pm|g|j,\bar m,\pm\rangle.
\ee
One finds that the $PSL(2,\Rbar)$ group manifold splits into six
regions where these wavefunctions are smooth. Generically, the
wavefunctions are not smooth at the boundaries separating these
regions. However, they are still globally defined.
When studying $PSL(2,\Rbar)$ by itself, one would usually prefer to
work with a basis of smooth wavefunctions.
\item
The coset conformal field theory $PSL(2,\Rbar)_{k<0}/U(1)$, where $k$
is the level of the $PSL(2,\Rbar)$ affine Lie algebra and $U(1)$ acts
as $g\mapsto e^{\alpha\sigma}ge^{\alpha\sigma}$, describes the
generalized Milne spacetime \eq{backgroundIIA} without the identification
\eq{uvid} \cite{Kounnas:1992wc,Craps:2002ii}.
Its globally defined wavefunctions are obtained by
restricting to those wavefunctions \eq{KSL} that are invariant under
$U(1)$, that is, to those satisfying $m=-\bar m$
\cite{Dijkgraaf:1991ba,Craps:2002ii}.
\item
The $\Zbar$ orbifold group is another (discrete) subgroup of
$U(1)_L\times U(1)_R$. The globally defined
orbifold wavefunctions are obtained by
imposing a quantization condition on $m$ (momentum quantization) and
by introducing twisted sectors (winding modes). We refer to
\cite{Craps:2002ii} for details. In the present paper, we are
interested in modes with no momentum or winding along the Milne circle.
\end{enumerate}


\section*{Appendix B: More on Global Wavefunctions}

In this Appendix, we provide some technical details on the global wavefunctions
$K_{\pm\pm,\vec{k}}$ used in Section~\ref{sec:string}.
For simplicity of notation, we will suppress the subscript $\vec{k}$.
As compared to \cite{Craps:2002ii,
Elitzur:2002rt}, we restrict the discussion to the case of interest in this
paper, as explained at the beginning of subsection~\ref{subsub:global}. In the
notation of \cite{Craps:2002ii}, this means that
\be\label{restr}
\lambda=\mu=-j~,
\ee
with $j$ defined in \eq{eq:43} and \eq{eq:50} as
\be\label{eq:50bis}
j=-{1\over 2}+i{E_{\vec{k}}\over2Q}, \ \ \  E_{\vec k}=\sqrt{\vec k^2-Q^2}.
\ee

We start by recalling some properties of the hypergeometric function
$F(a,b;c;z)\equiv _2F_1(a,b;c;z)$ \cite{Abramowitz}. Using the notation
\be\label{an}
(a)_n={\G(a+n)\over\G(a)},
\ee
this function has the power series expansion
\be\label{powerseries}
F(a,b;c;z)=\sum_{n=0}^\infty{(a)_n(b)_n\over(c)_n}{z^n\over n!}
,
\ee
which converges at least for $|z|<1$. It is useful to note that as $z\rightarrow 0$,
\begin{equation}
F(a,b;c;z)\longrightarrow 1
\label{eq:AA}
\end{equation}
for all values of the parameters $a$, $b$ and $c$.
The behavior for large $|z|$ can be obtained, for $b-a$ not integer, using the transformation formula
\beqa\label{transfform}
F(a,b;c;z)&=&{\G(c)\G(b-a)\over\G(b)\G(c-a)}(-z)^{-a}F(a,1-c+a;1-b+a;{1\over z})\cr
&&+{\G(c)\G(a-b)\over\G(a)\G(c-b)}(-z)^{-b}F(b,1-c+b;1-a+b;{1\over z}), \ \ \
\eeqa
which is valid for $|\arg(-z)|<\pi$.
Equation \eq{transfform} is singular for integer $b-a$. For $a=b$, it is replaced by
\beqa\label{transfformaa}
&&F(a,a;c;z)={\G(c)\over\G(a)\G(c-a)}(-z)^{-a}\sum_{n=0}^\infty
{(a)_n(1-c+a)_n \over (n!)^2}z^{-n}\cr
&&\ \ \ \ \ \ \ \ \ \ \ \ \  \times\left(\log(-z)+2\Psi(n+1)-\Psi(a+n)-\Psi(c-a-n)\right),\ \
\eeqa
where
\be\label{Psi}
\Psi(z)={\G'(z)\over\G(z)}.
\ee
Expression~\eq{transfformaa} is valid for $|\arg(-z)|<\pi$, $|z|>1$ and $c-a$ not integer.
Other transformation formulas we will use are
\be\label{transfformbis}
F(a,b;c;z)=(1-z)^{-a}F\left(a,c-b;c;{z\over z-1}\right)
\ee
and
\beqa\label{transfformtris}
&&F(a,b;a+b;z)={\G(a+b)\over\G(a)\G(b)}\sum_{n=0}^\infty{(a)_n(b)_n\over(n!)^2}(1-z)^n\cr
&&\ \ \ \ \ \ \ \ \ \ \ \ \times \left(-\log(1-z)+2\Psi(n+1)-\Psi(a+n)-\Psi(b+n)\right),\ \
\eeqa
the latter of which is valid for $|\arg(1-z)|<\pi$ and $|1-z|<1$. Also, recall that the Euler Beta function $B(a,b)$ is defined by
\begin{equation}
B(a,b)=\frac{\Gamma(a)\Gamma(b)}{\Gamma(a+b)}.
\label{eq:new1}
\end{equation}

In terms of these functions, the global wavefunctions $K_{\pm\pm}$ are defined as follows.\\

\noindent Region I:
\beqa\label{KI}
K_{++}(uv)&=&{1\over2\pi i}B(-j,-j)(-uv)^j F\left(-j,-j;-2j;{1\over uv}\right)\cr
K_{--}(uv)&=&{1\over2\pi i}B(j+1,j+1)(-uv)^{-j-1}F\left(j+1,j+1;2j+2;{1\over uv}\right)\cr
K_{+-}(uv)&=&0\cr
K_{-+}(uv)&=&{1\over\pi i}B(-j,j+1)F(-j,j+1;1;uv)
\eeqa
Region II:
\beqa\label{KII}
K_{++}(uv)&=&{1\over2\pi i}B(j+1,j+1)(-uv)^{-j-1}F\left(j+1,j+1;2j+2;{1\over uv}\right)\cr
K_{--}(uv)&=&{1\over2\pi i}B(-j,-j)(-uv)^j F\left(-j,-j;-2j;{1\over uv}\right)\cr
K_{+-}(uv)&=&{1\over\pi i}B(-j,j+1)F(-j,j+1;1;uv)\cr
K_{-+}(uv)&=&0
\eeqa
Region III:
\beqa\label{KIII}
K_{++}(uv)&=&{1\over2\pi i}B(-j,j+1)(uv)^j F\left(-j,-j;1;1-{1\over uv}\right)\cr
K_{--}(uv)&=&{1\over2\pi i}B(-j,j+1)(uv)^{-j-1}F\left(j+1,j+1;1;1-{1\over uv}\right)\cr
K_{+-}(uv)&=&{1\over2\pi i}B(-j,j+1)(1-uv)^j F\left(-j,-j;1;{uv\over uv-1}\right)\cr
K_{-+}(uv)&=&{1\over2\pi i}B(-j,j+1)(1-uv)^{-j-1}F\left(j+1,j+1;1;{uv\over uv-1}\right)\ \ \ \ \ \ \
\eeqa
Region IV:
\beqa\label{KIV}
K_{++}(uv)&=&{1\over2\pi i}B(-j,j+1)(uv)^{-j-1}F\left(j+1,j+1;1;1-{1\over uv}\right)\cr
K_{--}(uv)&=&{1\over2\pi i}B(-j,j+1)(uv)^j F\left(-j,-j;1;1-{1\over uv}\right)\cr
K_{+-}(uv)&=&{1\over2\pi i}B(-j,j+1)(1-uv)^{-j-1}F\left(j+1,j+1;1;{uv\over
uv-1}\right)\cr
K_{-+}(uv)&=&{1\over2\pi i}B(-j,j+1)(1-uv)^j F\left(-j,-j;1;1;{uv\over
uv-1}\right)
\eeqa
Region V:
\beqa\label{KV}
K_{++}(uv)&=&{1\over\pi i}B(-j,j+1)F(-j,j+1;1;1-uv)\cr
K_{--}(uv)&=&0\cr
K_{+-}(uv)&=&{1\over2\pi i}B(j+1,j+1)(uv-1)^{-j-1}F\left(j+1,j+1;2j+2;{1\over 1-uv}\right)\cr
K_{-+}(uv)&=&{1\over2\pi i}B(-j,-j)(uv-1)^j F\left(-j,-j;-2j;{1\over 1-uv}\right)
\eeqa
Region VI:
\beqa\label{KVI}
K_{++}(uv)&=&0\cr
K_{--}(uv)&=&{1\over\pi i}B(-j,j+1)F(-j,j+1;1;1-uv)\cr
K_{+-}(uv)&=&{1\over2\pi i}B(-j,-j)(uv-1)^j F\left(-j,-j;-2j;{1\over 1-uv}\right)\cr
K_{-+}(uv)&=&{1\over2\pi i}B(j+1,j+1)(uv-1)^{-j-1}F\left(j+1,j+1;2j+2;
{1\over 1-uv}\right)\cr &&
\eeqa

First, we use these explicit formulas to determine the behavior of some of these
wavefunctions in the asymptotic early and late time regimes. In
regions I and II, we can use \eq{uvtx} to write
\be\label{uvt}
uv=-\sinh^2(Qt),
\ee
while in regions V and VI
\be\label{uvtbis}
1-uv=-\sinh^2(Qt).
\ee
Let us concentrate on the early time part of Region I ($t\ll-1/Q$). It follows from
\eq{KI}, \eq{uvt} and \eq{powerseries} that as $t\to-\infty$
\be\label{K++early}
K_{++}\rightarrow {1\over2\pi i}B(-j,-j)2^{-2j}e^{-2jQt}=
{1\over2\pi i}B(-j,-j)4^{-j}e^{Qt}e^{-iE_{\vec{k}}t}.
\ee
This shows that $K_{++}$ reduces to a positive frequency wave for early times in
Region
I. Moreover, expression~\eq{K++early} allows us to determine the factor
 $\N$ such that
\be\label{calKnorm}
\K_{++}=\N K_{++}
\ee
is normalized (see subsection~\ref{subsub:global}). The factor $\N$ is actually
the same for all $K_{\pm\pm}$. Comparing \eq{K++early} with \eq{eq:48} or \eq{eq:49},
we find
\be\label{calN}
\N={4^je^{\phi_0}2\pi i\over \sqrt {E_{\vec{k}}} B(-j,-j)}.
\ee
Using \eq{transfform}, we can rewrite $K_{-+}$ in Region I as
\beqa\label{K-+I}
K_{-+}&=&{1\over \pi i}B(-j,2j+1)(-uv)^jF\left(-j,-j;-2j;{1\over uv}\right)\cr
      &&+{1\over \pi i}B(j+1,-2j-1)(-uv)^{-j-1}F\left(j+1,j+1;2j+2;{1\over uv}\right).\ \ \ \ \ \ \
\eeqa
This shows that asymptotically in Region~I, $K_{-+}$ is a superposition of positive
and negative frequency waves with equal amplitudes.

Equivalently to \eq{K-+I}, we can rewrite $K_{+-}$ in Region II as
\beqa\label{K+-II}
K_{+-}&=&{1\over \pi i}B(-j,2j+1)(-uv)^jF\left(-j,-j;-2j;{1\over uv}\right)\cr
      &&+{1\over \pi i}B(j+1,-2j-1)(-uv)^{-j-1}F\left(j+1,j+1;2j+2;{1\over uv}\right),\ \ \ \ \ \ \
\eeqa
which shows that asymptotically in Region~II, $K_{+-}$ is a superposition of positive
and negative frequency waves with equal amplitudes.

Next, we discuss the behavior near the Big Crunch/Big Bang singularity $uv=0$. First
consider $K_{++}$. In Region~I,
we can use \eq{transfformaa} and \eq{KI} to write
\be\label{K++Izero}
K_{++}(uv)={1\over 2\pi i}\sum_{n=0}^\infty{(-j)_n(j+1)_n\over(n!)^2}(uv)^n(-\log(-uv)
+2\Psi(n+1)-2\Psi(-j+n)).
\ee
We see that in Region I, $K_{++}$ diverges logarithmically near $uv=0$.
Similarly, we find in Region~II
\be\label{K++IIzero}
K_{++}(uv)={1\over 2\pi i}\sum_{n=0}^\infty{(-j)_n(j+1)_n\over(n!)^2}(uv)^n(-\log(-uv)
+2\Psi(n+1)-2\Psi(j+1+n)).
\ee
Applying \eq{transfformbis} and \eq{transfformtris} to \eq{KIII}, we find that
in Region~III
\beqa\label{K++IIIzero}
&&K_{++}(uv)={1\over 2\pi i}\sum_{n=0}^\infty{(-j)_n(j+1)_n\over(n!)^2}(uv)^n\cr
&&\ \ \ \ \ \ \ \ \ \ \ \  \ \times(-\log(uv)
+2\Psi(n+1)-\Psi(-j+n)-\Psi(j+1+n)).\ \ \ \ \ \ \
\eeqa
The same equation, \eq{K++IIIzero}, holds in Region~IV as well.
Now consider $K_{+-}$
near $uv=0$. In Region~I, it vanishes. The small $uv$ behavior in Region~II is manifest
from \eq{powerseries} and \eq{KII}. In Region~III, we can use \eq{transfformbis} to
rewrite \eq{KIII} as
\be\label{K+-IIIzero}
K_{+-}(uv)={1\over2\pi i}B(-j,j+1)F(-j,j+1;1;uv).
\ee
The same formula, \eq{K+-IIIzero}, also holds in Region~IV.
Therefore, as $uv$ approaches zero, $K_{+-}$ approaches a constant in regions
III and IV, and twice that constant in
Region~II.

\section*{Appendix C: Comparison with the Milne Orbifold}

In subsection~\ref{sub:string:fluctuations} we gave a qualitative discussion of some
differences and similarities between the pure and generalized Milne orbifolds. This
comparison can be made very concrete at the level of global wavefunctions, which we will do
in this Appendix. Some of these observations were first made by the authors of
\cite{Berkooz:2002je}. As in the previous Appendix, we will suppress the
subscript $\vec{k}$ for notational simplicity.

In the limit that
\be\label{limit}
uv\to0 \ \ {\rm with} \quad {\vec k^2uv\over Q^2}\ \ {\rm and} \quad Q r_{0} \quad {\rm fixed},
\ee
where $r_{0}$ is defined in \eq{uvid}, the equation of motion \eq{eq:37}
reduces to that of a massless scalar in the Milne orbifold \eq{eommilne}. We now
take this limit of the solutions $\K_{++}$ and
$\K_{+-}$ and see what they correspond to in the pure Milne case.
First, note that in the limit \eq{limit},
\be\label{limitEQ}
{E_{\vec{k}}\over Q}\to+\infty.
\ee
It then follows from \eq{eq:50bis} that
\be\label{limitj}
j\to-{1\over 2}+i\infty.
\ee
In this limit, the normalization factor \eq{calN} becomes
\be\label{Nlimit}
|\N|\to{e^{\phi_0}\sqrt\pi\over\sqrt{8Q}}.
\ee
Using the fact that
\be\label{Psiinfty}
\Psi(z)\simeq\log(z)\ \ \ \ {\rm for}\ z\to\infty \ {\rm with}\
|\arg(z)|<\pi,
\ee
we find from \eq{K++Izero} that in regions I and II
\be\label{K++milne}
K_{++}\to{1\over 2\pi
i}\sum_{n=0}^\infty{1\over(n!)^2}\left({E_{\vec{k}}^2uv\over 4Q^2}\right)^n
\left(-\log\left(-{E_{\vec{k}}^2uv\over4Q^2}\right)+\pi i+2\Psi(n+1)\right)
\ee
and
\be\label{K++milneII}
K_{++}\to{1\over 2\pi
i}\sum_{n=0}^\infty{1\over(n!)^2}\left({E_{\vec{k}}^2uv\over 4Q^2}\right)^n
\left(-\log\left(-{E_{\vec{k}}^2uv\over4Q^2}\right)-\pi i+2\Psi(n+1)\right)
\ee
respectively.
Since in Region II $uv=-Q^2t^2$ in the limit \eq{limit},
the right hand side of \eq{K++milneII} is precisely proportional to the
Hankel function $H^{(2)}_0(Et)$, that is,
\be\label{K++milneHankel}
K_{++}\to-{1\over 2}H^{(2)}_0(E_{\vec{k}}t),
\ee
while \eq{K++milne} is its analytic continuation to Region I with
negative $t$ via the lower half $t$ plane~\cite{Tolley:2002cv}.
Therefore, the limit \eq{limit} of $\K_{++}$ in regions I and II is the
wavefunction used in~\cite{Berkooz:2002je,  Tolley:2002cv} and~\cite{Nekrasov:2002kf} to define the adiabatic
vacuum inherited from Minkowski space.
In Region III, using \eq{K++IIIzero} and \eq{limitj}, we find that
\be\label{K++milneIII}
K_{++}\to{1\over 2\pi
i}\sum_{n=0}^\infty{1\over(n!)^2}\left({E_{\vec{k}}^2uv\over 4Q^2}\right)^n
\left(-\log\left({E_{\vec{k}}^2uv\over4Q^2}\right)+2\Psi(n+1)\right),
\ee
which is the expansion of the modified Bessel function $K_0$:
\be\label{K++milnemodHankel}
K_{++}\to{1\over \pi i}K_0\left(\sqrt{uvE_{\vec{k}}^2\over Q^2}\right).
\ee
The function $K_0$ decays asymptotically (that is, for large values of its argument).
This is exactly the analytic continuation of \eq{K++milneHankel} to
Region III and, additionally, to Region IV, via the lower half $t$ plane.
We conclude that in all four regions of pure Milne space I, II, III and IV, the limit \eq{limit}
of $\K_{++}$ exactly coincides with the wavefunctions used in
\cite{Berkooz:2002je} to define the vacuum inherited from Minkowski space.
These wavefunctions coincide with those of \cite{Tolley:2002cv} when
restricted to regions I and II.

To determine the behavior of $\K_{+-}$ , note that in the limit~\eq{limit}
\be\label{limithyper}
F(-j,j+1;1;uv)\to\sum_{n=0}^\infty{1\over (n!)^2}\left({uvE_{\vec{k}}^2\over4Q^2}\right)^n.
\ee
It follows that in Region II
\be\label{limithypertwo}
F(-j,j+1;1;uv)\to J_0(E_{\vec{k}}t),
\ee
with $J_0$ a Bessel function,
whereas in regions III and IV
\be\label{limithyperthree}
F(-j,j+1;1;uv)\to I_0\left(\sqrt{uvE_{\vec{k}}^2\over Q^2}\right),
\ee
with $I_0$ a modified Bessel function which has the property that it
grows asymptotically. For this reason, $I_0$ it is usually not considered in
discussions of the Milne orbifold, as we mentioned in
subsection~\ref{sub:string:fluctuations}. In the generalized Milne orbifold, there is
nothing wrong with growing behavior near $uv=0$ since regions III and IV do
not extend to infinity. In the limit \eq{limit}, where regions III and IV are
blown up to infinite size, the prefactor $B(-j,j+1)$ in \eq{KII} and \eq{K+-IIIzero}
actually goes to zero. Therefore, although in regions III and IV $K_{+-}$ is proportional
to \eq{limithyperthree} in the limit \eq{limit}, the proportionality factor is zero
and $K_{+-}$ vanishes, as it does in regions I and II.



\end{document}